\RequirePackage{fix-cm}
\documentclass[smallextended]{svjour3}       
\smartqed  
\usepackage{graphicx}
\usepackage{color}
\usepackage{amssymb}
\usepackage{amsmath}
\begin{document}

\title{Curvature effects in turbulent premixed flames of H$_2$/Air: a DNS study with reduced chemistry}
\titlerunning{Curvature effects in turbulent premixed flames of H$_2$/Air}        

\author{G. Rocco  \and  F. Battista \and F. Picano \and G. Troiani  \and C.M. Casciola}


\institute{
           G. Rocco \at
           Department of Aeronautics, Imperial College London, 
           South Kensington Campus, London SE7 2AZ, UK
     \and
            F. Battista \and C.M. Casciola \at
            Department of Mechanical and Aerospace Engineering,
            La Sapienza University, \\via Eudossiana 18, 00184, Roma, Italy
           \and
           F. Picano \at
           Department of Industrial Engineering, University of Padova,
           \\ Via Venezia 1, 35131 Padua, Italy
            \email{francesco.picano@unipd.it}
           \and
           G. Troiani \at
           Sustainable Combustion Laboratory, ENEA C.R. Casaccia, 
           via Anguillarese 301, 00123 Roma, Italy
            }

\date{Received: date / Accepted: date}
\journalname{Flow, Turbulence and Combustion}

\maketitle
\sloppy

\begin{abstract}
Data from a three-dimensional Direct Numerical Simulation of a 
turbulent premixed Bunsen flame at a low global Lewis number are analyzed to
address the effects of the curvature on the local flame front.
For this purpose, the chemical kinetics is modeled according to a
reduced scheme, involving 5 reactions and 7 species, to mimic
a H$_{2}$/Air flame at equivalence ratio $\phi=0.5$.
An increase of the local temperature and reaction rate is found 
for fronts elongated into the fresh gases (concave), while local quenching 
is observed for fronts elongated in the opposite direction (convex), i.e.\ 
towards the burnt mixture.
Data show that the occurrence in the reaction region of these super-reactive (concave fronts)
 and quenched zones (convex fronts)  is predominant compared to 
 a behavior compatible with the corresponding unstretched laminar flame.
In particular, well inside the reaction region, 
 the probability density function of the OH radical concentration shows a bi-modal shape 
 with peaks corresponding to
 negative (concave) and positive (convex)
 curvatures, while a locally flat front is less frequently detected. The two states are 
 associated with a higher and lower chemical activity with respect the laminar case.
Additional statistics conditioned to the local hydrogen concentration provide
further information on this dual-state dynamics and on the differences
with respect to the corresponding laminar unstretched flame when moving from the fresh to the 
burnt gas regions.
{
Finally we discuss the effects of the turbulence on the thermo-diffusive instability 
showing that the turbulent fluctuations, increasing the flame front corrugations, are 
essentially responsible of the local flame quenching.
}

\keywords{DNS\and turbulent premixed flames \and Hydrogen \and flame instability}
\end{abstract}

\section{Introduction}
\label{sec:intro}
An increasing concern for the environment and the rising cost of fossil fuels are
driving modern designers to consider different types of fuels for power 
production or  transport systems. Hydrogen and 
hydrogen-based fuel mixtures, which contain methane, carbon monoxide or 
small traces of fossil gases (the so-called \emph{syngas}~\cite{syngas}) are among
the most promising fuels for applications in the near future.
The key features of hydrogen-based fuels are their 
low polluting emissions, their high reaction heat values and their abundance in the 
natural environment.

One of the most distinctive features of gaseous hydrogen is its low molecular mass in 
comparison with that of carbon-bases fuels, 
which is responsible for its faster diffusion speed. 
The high mass diffusion of hydrogen strongly influences mixing and combustion 
processes, and induces the thermo-diffusive instabilities, which may cause local flame quenching 
and high temperature fluctuations~\cite{aspdaybel,chacan,gicque2005influence,vreoijgoebas,wanluofan,wanluoqiulufan}.
Within the framework of combustion theory, the effects of mass diffusion  are properly described by 
the Lewis number~\cite{chacan,chulaw,lawchu,lipatnikov1998lewis,Lipatnikov20051,yuajulaw}, namely the 
ratio between the heat and the mass diffusion coefficients, which properly 
summarizes the instability characterizing the flame behavior. 
When a mixture is composed of several species with different transport coefficients, 
the global Lewis number is assumed to be equal to the Lewis number of the less abundant 
species in the mixture, evaluated with respect to the stoichiometric balance~\cite{poivey}.
When the Lewis number is much lower than one, thermo-diffusive instabilities occur; 
there are consequent quenching effects in the local flame and the appearance 
of regions where the reactions peak. This phenomenon has been 
properly explained in the context of a laminar flame \cite{law}. 

Using a simplified chemistry, Chakraborty et al.~\cite{chacan} studied the effects of the global Lewis 
number on the scalar transport properties for some turbulent premixed flames.
Three-dimensional Direct Numerical Simulation (DNS) of the statistically 
planar flames were performed in a wide range of different Lewis numbers, above and below the critical value $Le=1$.
In these simulations, the flames with ${\rm Le}\ll 1$ exhibited strong counter-gradient 
transport effects, while the gradient transport (Fick-like law) was detected when the Lewis number 
was increased. 
Here couter-gradient transport is taken to imply $J_D \cdot \nabla c > 0$, as opposed to Fick-like transport where $J_D \cdot \nabla c < 0$, with $J_D$ the diffusive flux and c the concentration.
Although Chakraborty et al. have addressed the effects  of a properly
defined global Lewis number on flames in realistic 
configurations, the local effects due to the  different mass 
diffusivity of the chemical species was not considered.  

A similar configuration, though two-dimensional,  is used in Chakraborty et 
al.~\cite{chahawchecan} to address the correlation between the strain rate and 
the curvature, and their effects on the surface density function. Their analysis takes into account two turbulent premixed flames of 
lean methane-air and hydrogen-air mixtures. Data are obtained by means of Direct 
Numerical Simulations and a suitably detailed scheme is used to model the chemistry. The authors
compare the statistics of methane-air and hydrogen-air 
flames and investigate the effects of thermochemistry and preferential 
diffusion on the surface density function. The surface 
density function, $\sigma=\left|\nabla c\right|$ is an important observable 
in turbulent premixed flames since it is strictly correlated to the dissipation 
rate, $\chi=D\sigma^2$, which is crucial in turbulent combustion 
modeling~\cite{manbor}.

A similar geometry is investigated in~\cite{shitantanmiy}, where DNS of 
hydrogen-air turbulent premixed flames is used  to investigate 
the local flame structure and the fractal characteristics. The authors show the 
strong dependence of the flame structure on the Reynolds number: local 
quenching is seen to increase with the Reynolds number.

Im \& Chen~\cite{imche} discuss  the interaction of premixed flames 
with two dimensional turbulence by performing two numerical simulations.
They compare a lean and a rich hydrogen/air premixed flame, showing the
opposite effect of the curvature and strain on the chemical activity of the two flames.
The effects of thermo-diffusive instability on the initially planar  two dimensional hydrogen 
flame is investigated in~\cite{basvre}. The authors perform the 1D and 2D simulations of a lean 
premixed hydrogen flame, showing that even a 2D initially planar flame is unstable, self inducing
 flame wrinkling and fluctuations.
A wider analysis on the statistically planar turbulent flame in 3D is reported 
in~\cite{aspdaybel,daybelbrepasbec}. The authors address the behavior of a lean premixed
hydrogen flame reporting the effects of Karlovitz number, Ka, and 
equivalence ratio, $\phi$, on the flame instability (the cellular flame regime).
They observe that at  high Karlovitz number the distributed flame 
regime occurs showing a broad flame front which is 
similar to a mixing region.

Wang et al.~\cite{wanluofan} have investigated the effects of a swirling flame 
on the preferential diffusion in hydrogen-air flames, showing that the 
swirl tends to suppress the effects of the preferential diffusion. 

The purpose of the present study is to investigate the effects of the preferential 
diffusion of hydrogen on a Bunsen turbulent premixed air flame focusing on the effect
of turbulent fluctuations on the local flame dynamics. 
The hydrogen/air mixture is
provided by a fully developed turbulent pipe flow obtained by a companion simulation which
carries \emph{real} turbulent fluctuations at the Bunsen inlet.
The present geometry, being well replicable in experiments, represents a
step forward in direct numerical simulations of turbulent combustion.

{
Consistently with previous studies~\cite{imche,aspdaybel,daybelbrepasbec},  
we  found the typical cusp-like structures of the flame front protruding towards 
products associated with quenched regions, while opposite structures induce super-burning regions.
We statistically characterize this behavior in the turbulent regime
highlighting  the differences among the local turbulent flame structure, the unstrained 
laminar flame and the laminar (cellular) flame subjected only to thermo-diffusive instability.
To this end, we provide joint-pdf among local flame front curvature,
local velocity gradient, concentration of OH radicals, and 
hydrogen atomic equivalence ratio with respect to laminar cases.}

All the statistics are conditioned to different intervals
of the local hydrogen concentration allowing a study of the process from the beginning to the end of the 
reaction region and a direct comparison with the  unstretched laminar flame. Well inside the local flame, we show that the dynamics
strongly differs from  
the corresponding unstrained premixed flame and two different burning states are typical.
One is a super-burning cell with high chemical activity 
and temperature higher than  adiabatic;  the other corresponds to an 
almost quenched region. The local state is mainly determined by the local curvature of the flame front.
{A comparison between the turbulent and
the corresponding laminar Bunsen  flame is provided to assess the effects of turbulence on the front dynamics. 
Although the cellular structure induced by the thermo-diffusive instability of the laminar flame presents super-burning 
regions resembling those observed in the turbulent flame, quenched regions are considerably less frequent.
%
}

The paper is organized as follows. Section \S~\ref{sec:nummet}
describes the numerical tool employed for the Direct Numerical Simulation
of the 3D turbulent flame, the chemical mechanism
and the used physical parameters;
section \S~\ref{sec:results} discusses the results obtained from DNS: 
instantaneous analysis and statistical analyses based on the joint probability density function are used;
final remarks and comments are illustrated in the last section \S~\ref{sec:concl}.

\section{Numerical methodology}
\label{sec:nummet}
\subsection{Flow solver}

The problem is described by the Navier-Stokes equations in cylindrical coordinates in an open
environment at constant pressure $p_0$. The governing equations are expanded by means of a 
low-Mach number asymptotic approximation~\cite{majset,battista2014turbulent},
which, in dimensionless form, read:

\begin{align}
\label{eq:cont_def}
&\frac{\partial \rho}{ \partial t} + \nabla \cdot (\rho {\bf u}) = 0  
\\
\label{eq:mom_def}
&\frac{ \partial \rho \bf u}{\partial  t} + \nabla \cdot ({\rho \bf u \otimes \bf u}) =
\frac{1}{\rm Re}\nabla \cdot {\Sigma} - \nabla P + \rho {\bf g^*} \\
\label{eq:spec_def}
&\frac{\partial \rho Y_a}{\partial  t} + \nabla \cdot ({\rho {\bf u }Y_a}) =
\frac{1}{{\rm Re Sc}_a} \nabla \cdot ({\sigma_D} \nabla Y_a) + {\omega_a} \\
\label{eq:div_def}
&\nabla \cdot {\bf u} =\frac{1}{p_0}\left[ 
\frac{1}{\rm Re Pr } \nabla \cdot ({\sigma_T} \nabla T)
 + \frac{\gamma -1}{\gamma}{\omega_Q} \right]\\
\label{eq:state_def}
&T=\frac{p_0}{\rho}
\end{align}
where ${\bf \Sigma}=2\,{\sigma_\mu(T)}\,{\bf E} {+ \sigma_\lambda \,tr({\bf E}){\bf I}}
={\sigma_\mu(T)}\,(\nabla \bf u + {\nabla \bf u}^{T}) {+ \sigma_\lambda \,\nabla \cdot {\bf u}\,{\bf I}}$ 
is the viscous stress tensor 
with $\sigma_\lambda$ set to zero and $\sigma_\mu(T) = \mu/\mu_0$
denoting the dimensionless, temperature ($T$) dependent viscosity. $\rho$, $\bf u$, $p_0$
and $P$ are the density, the velocity, the thermodynamic pressure, and the dynamic pressure, respectively.  
We stress that the thermodynamic pressure $p_0$ is 
constant in space, due to the low-Mach number expansion~\cite{majset},
and in time, due to the open space conditions. The assumed dimensionless
  is $p_0=1$, corresponding to a pressure of 1 atm.
$Y_{a}$ and $\omega_{a}$  are the concentration and the global
reaction rate of the $a^{th}$ species in the mixture. 
$\sigma_D(T) = D_a/D_a^0$ and $\sigma_T(T) = \alpha/\alpha_0$ 
are the dimensionless mass and thermal diffusivities.
 The dynamic viscosity of the mixture 
at reference conditions (inlet conditions $p=1$atm and $T=300$K)
is indicated by $\mu_0$, the Reynolds number is  ${\rm Re}= \rho_0 {U_0 L_0}/\mu_0$, {where $U_0$ and
$L_0$ are the typical velocity and length scales (bulk velocity and radius of the jet in the present case)}.
${\rm Pr}=\mu_0/(\rho_0 \alpha_0)$ is the Prandtl number
at reference conditions.
The ratio of the constant 
pressure coefficient, $c_p$, and the constant volume coefficient, $c_v$,
is denoted by $\gamma$; the Schmidt number ${\rm Sc_a}$ for the $a^{th}$ species is defined as the  ratio of the viscosity and the
binary mass transport coefficient of the specific species $a$ in the most 
abundant element in the mixture (here taken to be  N$_2$): ${\rm Sc}_a=\mu_0/(\rho_0 {\cal D}_{a{\rm N}_2})$.
In these conditions, the Schmidt number only depends on the a-th specie and on the 
most abundant one (at reference conditions), see~\cite{law} sec. 5.2.4.1.
Details on the values of the parameters and the assumptions used in 
modeling the mass diffusion coefficients are given in section \ref{sec:kem_mod}.  

The Direct Numerical Simulation (DNS) of  
the Bunsen flame is performed using second order central finite differences in
 a conservative formulation on a staggered grid. Bounded Central Difference 
Schemes~\cite{watdec} are used for the non-linear terms in the scalar 
equations~\eqref{eq:cont_def} and~\eqref{eq:spec_def}.
Temporal evolution is performed by a low-storage third order 
Runge-Kutta scheme.  

\begin{figure}[h!]
\centering
  \includegraphics[width=.45\textwidth]{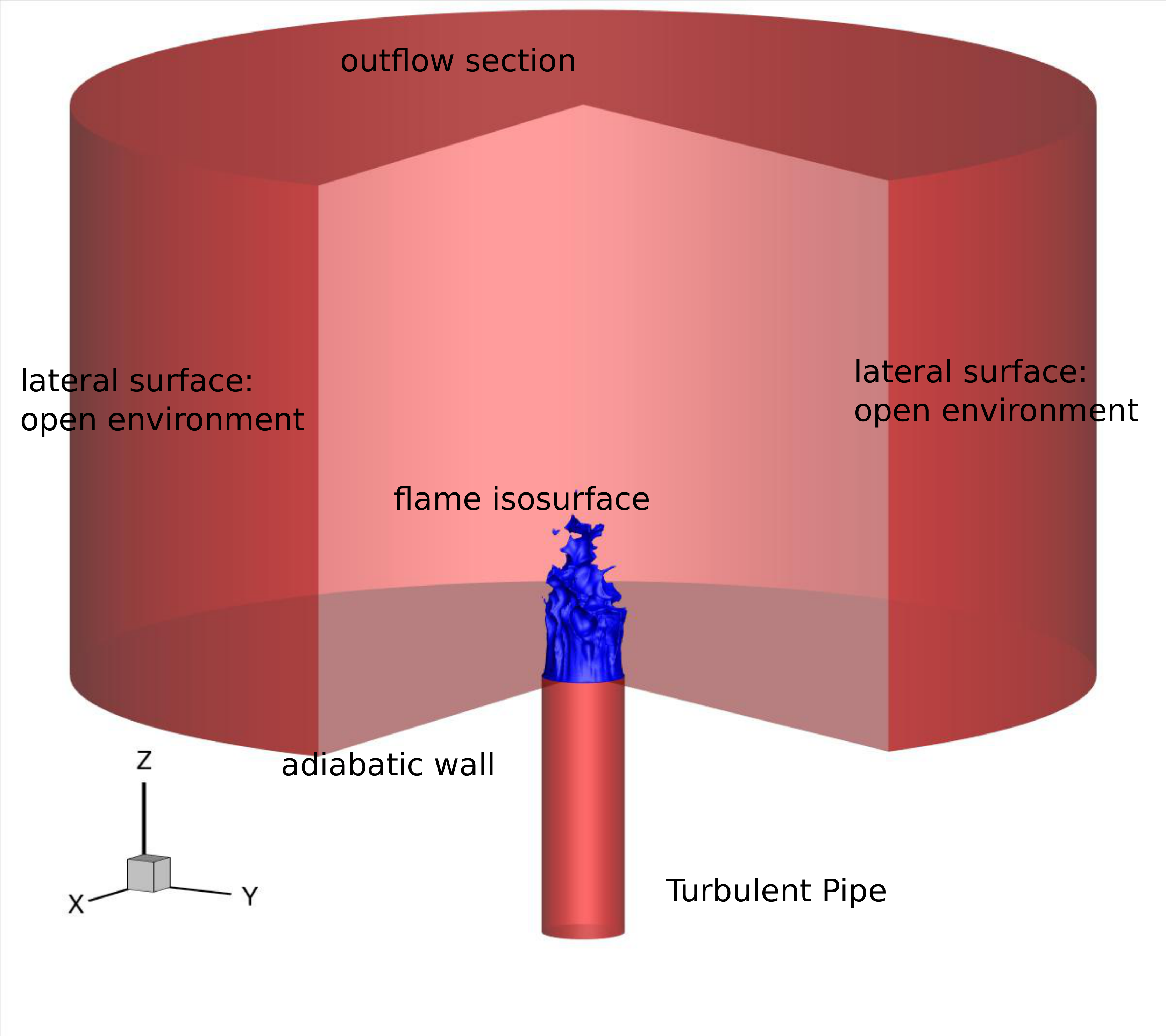}
\hspace{1.cm}
  \includegraphics[width=.3\textwidth]{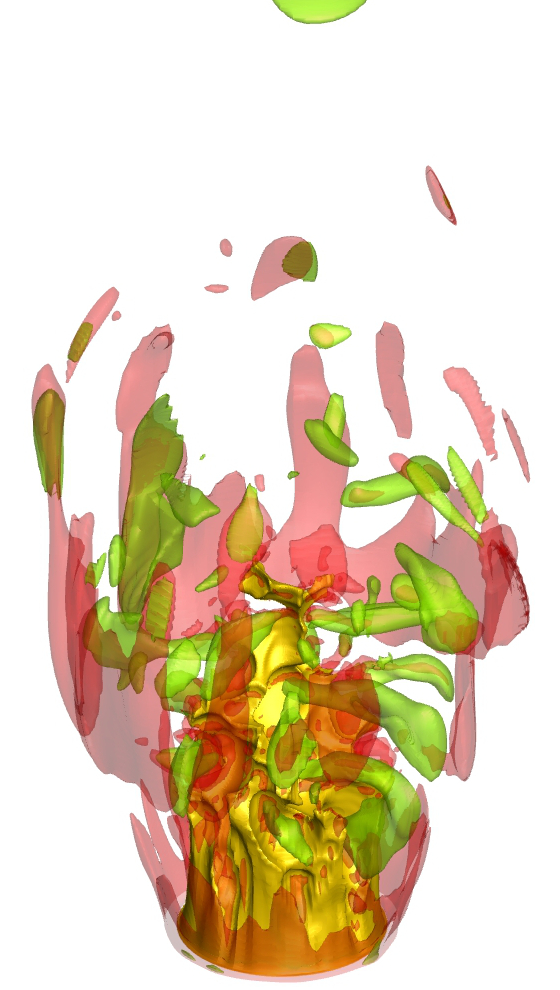}
{\scriptsize \put(-300,175){\bf (a)}}
{\scriptsize \put(-100,175){\bf (b)}}
\caption{\label{fig:3D} Panel (a) reports a sketch of the domain geometry with the 
isosurface (in blue) of the hydrogen mass fraction. Panel (b) reports an instantaneous
flame configuration. Yellow: hydrogen mass fraction isosurface, $Y_{\rm H_2}=0.01$; green:
NO mass fraction isosurface, $Y_{\rm NO}=5.9\times 10^{-8}$; red: water mass fraction 
isosurface $Y_{\rm H_2 O}=0.3$ }
\end{figure}

Time-dependent boundary conditions are prescribed for the inflow. 
For this purpose, a fully turbulent inflow velocity is dynamically 
assigned by means of a cross-sectional slice of a fully developed 
pipe flow simulation. Indeed, the inflow 
condition is obtained by a companion simulation of a fully turbulent periodic pipe
flow from which the turbulent velocity profile in a cross-sectional slice is extracted at each time step 
and  enforced at the inlet surface of the jet.
This allows a realistic turbulent inlet.
The computational domain consists 
of a cylinder: the flow is injected in one base and  streams out at the other end.
Injection occurs through an orifice in the impermeable 
and adiabatic base of the domain, see figure~\ref{fig:3D} panel (a).
The density and concentrations of the species
are constant and uniform along the inflow orifice. A convective 
Orlanski~\cite{orla,pic_cas} condition is adopted at the outflow section. 
The lateral surface of the cylinder is modeled by an adiabatic 
traction-free~\cite{danboe} condition, which makes the entrainment of
external fluid possible. 

The numerical integration starts knowing the density $\rho^n$, momentum 
$(\rho {\bf u})^n$, and species mass fraction $(\rho Y_a)^n$ fields at a generic time 
step $n$. We remark once more that, for low-Mach number expansion and the present open environment configuration, 
the thermodynamic pressure is constant both in space  
and in time. Hence given the density $\rho^n$, the temperature $T^n$ can be evaluated through the equation of state, 
eq. \eqref{eq:state_def}. The main lines 
of the solution algorithm are listed below,
\begin{itemize}
\item[-] The mass conservation equation is used to integrate the fluid density at 
time step $n+1$, $\rho^{n+1}$.
\item[-] The momentum equation (deprived of the hydrodynamic pressure term) 
is used to find the unprojected velocity field, $(\rho \bf v)^{n+1}$.
\item[-] The species continuity equation is used to advance in time $(\rho Y_a)^{n+1}$.
We remark that the source terms in species equations are only function of 
temperature, pressure, and species mass fractions at the time step $n$.
\item[-] The velocity is then projected enforcing the local value of the divergence which depends
on the diffusive temperature flux and combustion heat release. The projection method is the straightforward
extension of the classical Chorin's method usually employed for fully incompressible Navier-Stokes 
equations integration. This step requires the solution of a linear system which is solved using 
the SuperLU Library~\cite{superlu_ug99}.
\end{itemize}

The numerical code has been fully tested in several 
configurations ranging from cold incompressible jets, see 
e.g. \cite{pic_cas,pichan,picsarguacas}, to reactive premixed Bunsen flames, see e.g. 
\cite{batpictrocas,picbattrocas,battista2014turbulent}.

\subsection{Reduced chemical model}
\label{sec:kem_mod}
The chemical kinetics model is known to deeply affect the dynamics of the flame. Although global 
chemistry schemes facilitate the incorporation of complex chemical features into turbulent simulations, 
the absence of intermediate radicals is unsuitable for addressing instabilities and extinction 
phenomena. 
In this work, a reduction of the original mechanism included in GRI-mech 2.1~\cite{grimech} is 
employed. The reduced mechanism implemented is made by 5-step and  
includes 4 radicals (H, O, OH, NO) besides the main reactants (H$_2$, O$_2$, H$_2$O, 
N$_2$)~\cite{chechakos} and reads:
\smallskip
\begin{enumerate}\label{reduced2}
\item H+O $\rightleftharpoons$ OH
\item H+O$_2 \rightleftharpoons$ OH+O
\item H$_2$+OH $\rightleftharpoons$ H$_2$O+H
\item H$_2$+O $\rightleftharpoons$ H+OH
\item N$_2$+O$_2 \rightleftharpoons$ 2\,NO.
\end{enumerate}
The GRI-mech scheme includes results published in the most recent literature and has been validated in a wide range 
of physical conditions (temperature between 1000 and 2500 K, pressure from 10 Torr to 10 atm) and 
equivalence ratios ($\phi$ from 0.1 to 5 for  premixed systems). One of its advantages
is its optimized chemical kinetics, which includes, via a quasi-steady state and local equilibrium assumptions, the 
effects of a larger number of elementary reactions  on the target species. However, 
the quasi-steady state assumption concerning the kinetics of the H$_2$ O$_2$  and HO$_2$ might  lead 
to some inaccuracies in the ignition of the mixture at  moderate temperatures and high pressures ~\cite{law}.
This effect is not deemed crucial for the present work  since the assumed pressure and temperatures 
are $p=1\,atm$, $300\,K<T<1550\,K$ respectively  and the emphasis is on the statistically steady state rather 
than ignition.  The complete GRI-mech scheme includes helium, neglected in the present simulations 
since  its role is mainly confined to the generation of pollutants, a subject beyond the scope of this paper.
The formation enthalpies of the species $\Delta h_{f}^0$ have been chosen from the NASA-Lewis and 
the Technion archives.  Their values are used to compute the heat release 
$\omega_{Q}=-\sum_{a=1}^{N_s} \Delta h_{f,a}^0 \dot{\omega}_a$, where $N_s$ is the 
number of species and $\dot{\omega_a}$ their respective global production rate. 

Binary mass diffusion coefficients $\mathcal{D}_{jk}$ of species 
$j$ into species $k$ may have an impact on the flame structure, as well as  on the 
formation and evolution of thermo-diffusive instabilities. The exact evaluation of 
these coefficients has an enormous computational cost since it requires the resolution 
of an $N_s \times N_s$ linear system at each grid point and time instant. 
This results in a strong limitation for approaches like DNS, where
a huge number of grid  points and small time steps are needed 
to describe the finest scales of the turbulent flame. 
A viable alternative is Fick's diffusion, whose adoption is here motivated in terms of
the \emph{Hirschfelder and Curtiss approximation}~\cite{hircurbir}.   
The equivalent diffusion coefficient $D_a$ of a  single species into the rest of 
the mixture~\cite{poivey} is given by
\begin{equation}
D_a=\frac{1-Y_a}{\begin{matrix}\sum_{j \ne a} X_j/ \mathcal{D}_{aj} \end{matrix}}
\end{equation}
where $X_j$ is the mole fraction of the generic species $j$ and can be reduced with reasonable accuracy to the 
binary diffusivity of species $j$ into nitrogen,  $D_{a} \simeq \mathcal{D}_{a{\rm N_2}}$, due to the excess of N$_2$ in the air mixture,
see~\cite{law} section 5.2.4.1 or~\cite{poivey} section 1.1.5. 
The binary diffusion coefficients $\mathcal{D}_{a{\rm N_2}}$ can be evaluated 
using the assumption of binary collisions, which holds when the gas is sufficiently 
diluted, hence  ternary and higher order collisions are rare enough that  their effects can be neglected~\cite{law}.  
Clearly, as a counterpart to the substantial  reduction in computational complexity,
small errors in the diffusion should be tolerated near the flame front.  
Despite the strong simplification, the diffusivity still   strongly depends on the temperature. In the present case, the temperature dependence is 
accounted for fixing the Schmidt number, ${\rm Sc}_a=\mu_0/(\rho_0 \mathcal{D}_{a {\rm N}_{2}})$,
such that the mass diffusivities and the thermal
diffusivity inherit the temperature dependence of the dynamic viscosity 
$\sigma_\mu(T)=\sigma_a(T)= \sigma_T(T)= (T/T_0)^{0.5}$, which follows from a Sutherland-like law. The proportionality is 
fixed by evaluating  the diffusion coefficients at $298 \,K$ ( see table~\ref{tab:diff}).
\vspace{0.5cm}
\begin{table}[htp]
\begin{center}
\begin{tabular}{||c|c|c||} \hline
\textbf{Species}&$\mathcal{D}_{k,N_2} [m^2/s]$& $Sc_k$ \\ \hline \hline 
H$_2$  & $8.56  \cdot\ 10^{-5}$ & 0.175 \\ \hline
O$_2$  & $5.931 \cdot\ 10^{-6}$ & 2.53  \\ \hline
H$_2$O & $1.218 \cdot\ 10^{-5}$ & 1.232 \\ \hline
H      & $2.136 \cdot\ 10^{-4}$ & 0.13  \\ \hline
O      & $1.094 \cdot\ 10^{-4}$ & 0.134 \\ \hline
OH     & $3.308 \cdot\ 10^{-5}$ & 0.45  \\ \hline
NO     & $5.625 \cdot\ 10^{-6}$ & 2.66  \\ \hline
He     & $3.06  \cdot\ 10^{-6}$ & 0.49  \\ \hline
\end{tabular}
\caption{$\mathcal{D}_{k,{\rm N}_2}$ and ${\rm Sc}_a=\mu/(\rho \mathcal{D}_{a,{\rm N}_{2}})$ are
the binary diffusion coefficients and the Schmidt numbers evaluated for the $a$ specie with respect to molecular nitrogen N$_2$ at $298K$,
respectively.
\label{tab:diff}
}
\end{center}
\end{table} 

Due to the excess of N$_2$, the dependence of $\gamma$ on the composition of the mixture may 
also be safely neglected and, given the involved temperature range (burnt/unburnt gas 
temperature ratio $T_b/T_u\simeq 5$), 
the constant value $\gamma=1.25$  can be considered a reasonable
approximation~\cite{poivey}.
\begin{figure}[t!]
\centering
\includegraphics[width=0.75\linewidth]{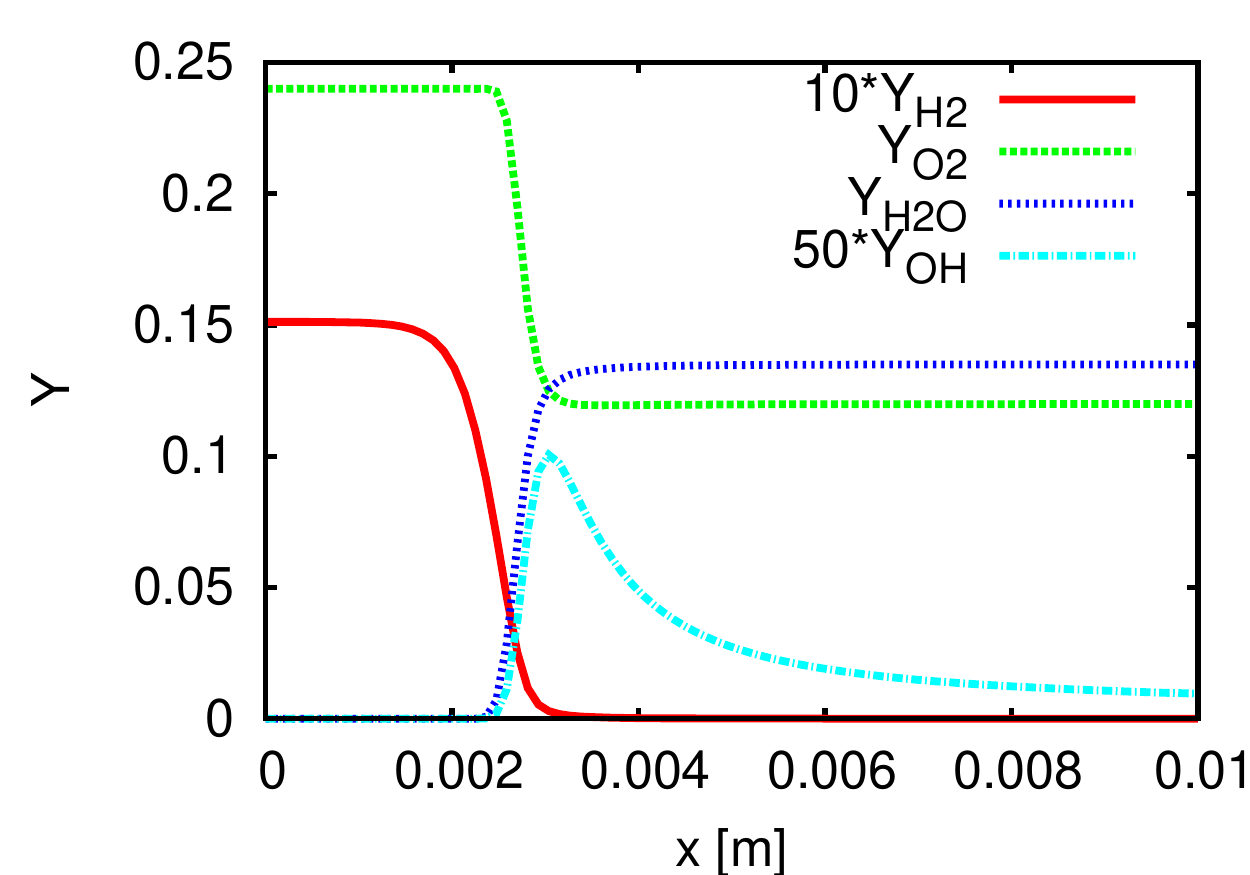}
\\[.05\linewidth]
\includegraphics[width=0.75\linewidth]{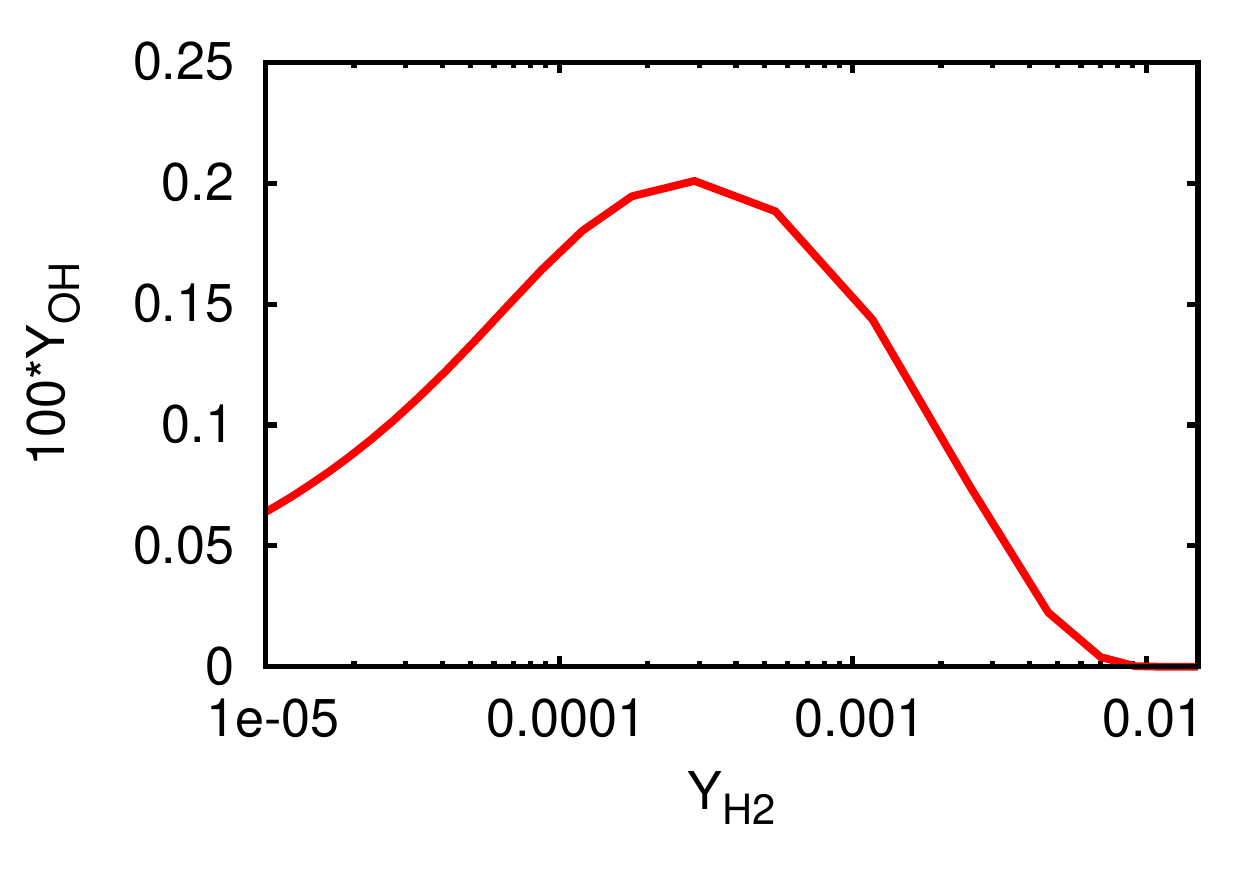}
\caption{\label{fig:0}
One-dimensional flame at $\Phi=0.5$.
Top panel, mass fractions $Y_k$ of reactants, products and OH radical vs dimensional distance.
Bottom panel, OH mass fraction $Y_{\rm OH}$ with respect to the mass fraction of molecular hydrogen $Y_{{\rm H}_2}$.
}
\end{figure}

Representative computational results for a one-dimensional, 
planar, and stationary flame at $\phi=0.5$ are reported in figure~\ref{fig:0}. 
The parameters of this preliminary simulation match the fully 3D ones that 
will be presented in the following section.
The adiabatic flame temperature $T_{ad}\simeq 1555\,K$ and the flame speed 
$S_L=0.35\,m/s$ are consistent with the available results in the literature, 
see e.g.~\cite{law}.
The top panel of figure~\ref{fig:0} represents the mass fractions as a function of the streamwise coordinate. 
At this equivalence ratio, hydrogen is completely consumed, while about $48\%$  of $O_2$ does not 
react. The narrow reaction zone spans the region extending from $x=0.002\,m$ to $x=0.0035\,m$ where
the radical concentrations are more prominent. The  concentration of the radical OH can be used to detect the reaction, as 
in experimental measures on H$_2$-lean flames \cite{chechakos}. The ability of the radical OH to mark the progress of the reaction
is confirmed by the bottom panel of figure \ref{fig:0} which shows the OH-mass fraction, $Y_{\rm OH}$, 
as a function of the H$_2$ mass fraction, $Y_{\rm H_2}$. This relation is valid for the one-dimensional laminar planar flames and will be referred as 
$Y_{\rm OH} = Y_{\rm OH}^L(Y_{{\rm H}_2})$.
When the reaction progresses, $H_2$ is consequently consumed and the OH mass fraction
abruptly increases and  reaches a peak at $Y_{{\rm H}_2}=2 \times 10^{-4}$ when all the fuel is almost burnt. 
Afterwards the OH mass fraction starts a more gradual decrease as a consequence of slow recombination phenomena. 

The reason for our detailed description of the one-dimensional laminar flame is related to the forthcoming use
of the local turbulent fluctuations of OH mass fraction $Y_{OH}(\mathbf{x}, t)$, evaluated with respect to the laminar case, to measure the local
activity of the combustion in the turbulent flame. The procedure is based on a statistical sampling of OH and ${\rm H}_2$ 
concentrations in the turbulent flame.  Given a generic turbulent event where the concentration of $OH$ and $H_2$ are  $Y_{\rm OH}^T$, $Y_{{\rm H}_2}^T$ respectively,
we associate the corresponding laminar state  $Y^L_{\rm OH}(Y_{{\rm H}_2}^T)$,
where the function $Y_{\rm OH} = Y^L_{\rm OH}(Y_{{\rm H}_2})$ is the profile of the $OH$ concentration in the laminar flame (see bottom panel of
figure~\ref{fig:0}). 

%
A positive/negative deviation with respect to the flamelet mass fraction (flamelet fluctuation),
\begin{equation}
	Y'_{\rm OH}=Y_{\rm OH}-Y^L_{\rm OH}|_{Y_{{\rm H}_2}}\, ,
\label{e:y'_oh}
\end{equation}
denotes more/less chemical activity with respect to the corresponding laminar state.
It is worth to remark that this observable allows to highlight the differences between the local
turbulent flame and the corresponding laminar unstretched flame. 
More specifically, the comparison is exerted by providing the 
deviation of the local turbulent quantities from the laminar ones at the corresponding state defined by
the same hydrogen concentration $Y_{{\rm H}_2}$.

\subsection{Direct Numerical Simulation details}
The parameters used in the DNS have been chosen to reproduce a  laboratory scale  
premixed Bunsen flame of a lean H$_2$-Air mixture. The fuel is issued from a pipe 
whose radius is $R=0.009\, \rm m$; the bulk velocity is $U_0=5\, \rm m/s$ and the equivalence 
ratio is  $\phi=0.5$. {The surrounding environment is filled by pure air at the same temperature 
of fresh gas, i.e.\ $T=298 K$.} The Reynolds number is 
${\rm Re}_{R}=U_{0}\,R/\nu_{\infty}=3000$ while the Prandtl number is
${\rm Pr} = \nu_0/\alpha_0=0.6$, where $\nu_0$ and 
$\alpha_0$ are the inflow kinematic viscosity and the thermal diffusivity
of the mixture, respectively. {The Froude number, taking into account the buoyancy 
effects, is fixed at $Fr = U_0 / \sqrt{g \, R}=16.8$}.
The Damk{\"o}hler and Karlovitz numbers at the inflow are 
$Da= L_t S_L/(u' \delta_L) \simeq 22.3$ and $Ka = \delta_L u_\eta /(\eta S_L)\simeq0.65$, respectively.
Here $L_t$ and $u'$ are the integral legth and the velocity root-mean-square (bulk) fluctuations, while $\eta$ and $u_\eta$ 
are the Kolmogorov length and velocity, respectively.
 The computational domain 
$[\theta_{max}\times R_{max} \times Z_{max}]=[2\pi \times 12\,R \times 14\,R]$ 
is discretized using a mesh of $N_\theta\times N_r\times N_z=128\times201\times560$ 
collocation points, endowed with a suitable radial stretching to properly capture the shear 
layers and the flame front dynamics. The characteristic grid spacing near the nozzle exit is 
$\Delta \simeq 2 \eta_{k}$, where $\eta_{k}$ is the Kolmogorov length scale near the wall 
of the adjoining pipe; the grid spacing was chosen such that at least four grid points are located across the 
instantaneous flame front. Additional details and comparisons with experimental data 
can be found in \cite{batpictrocas,picbattrocas}, where an identical numerical setup
is used to simulate a  Bunsen premixed flame using a global reaction model. 

{
Two 3D simulations have been performed and analyzed.  The former reproduces a turbulent Bunsen flame by 
enforcing a turbulent inflow velocity endowed with turbulent fluctuation extracted from a companion turbulent 
pipe simulation;
 the second one reproduces the laminar (cellular) Bunsen flame
by prescribing a parabolic velocity profile at the jet inlet. All other parameters 
are the same for the
two simulations to directly assess 
the effects induced by the turbulent fluctuations.}
The turbulent flame simulation spans about 4.5 $R/U_0$ time scales in the statistical steady 
state to collect about 60 uncorrelated fields every 0.075 $R/U_0$ to calculate the statistics.

\section{Results}
\label{sec:results}

\begin{figure}[t!]
\centering
\includegraphics[width=0.5\linewidth]{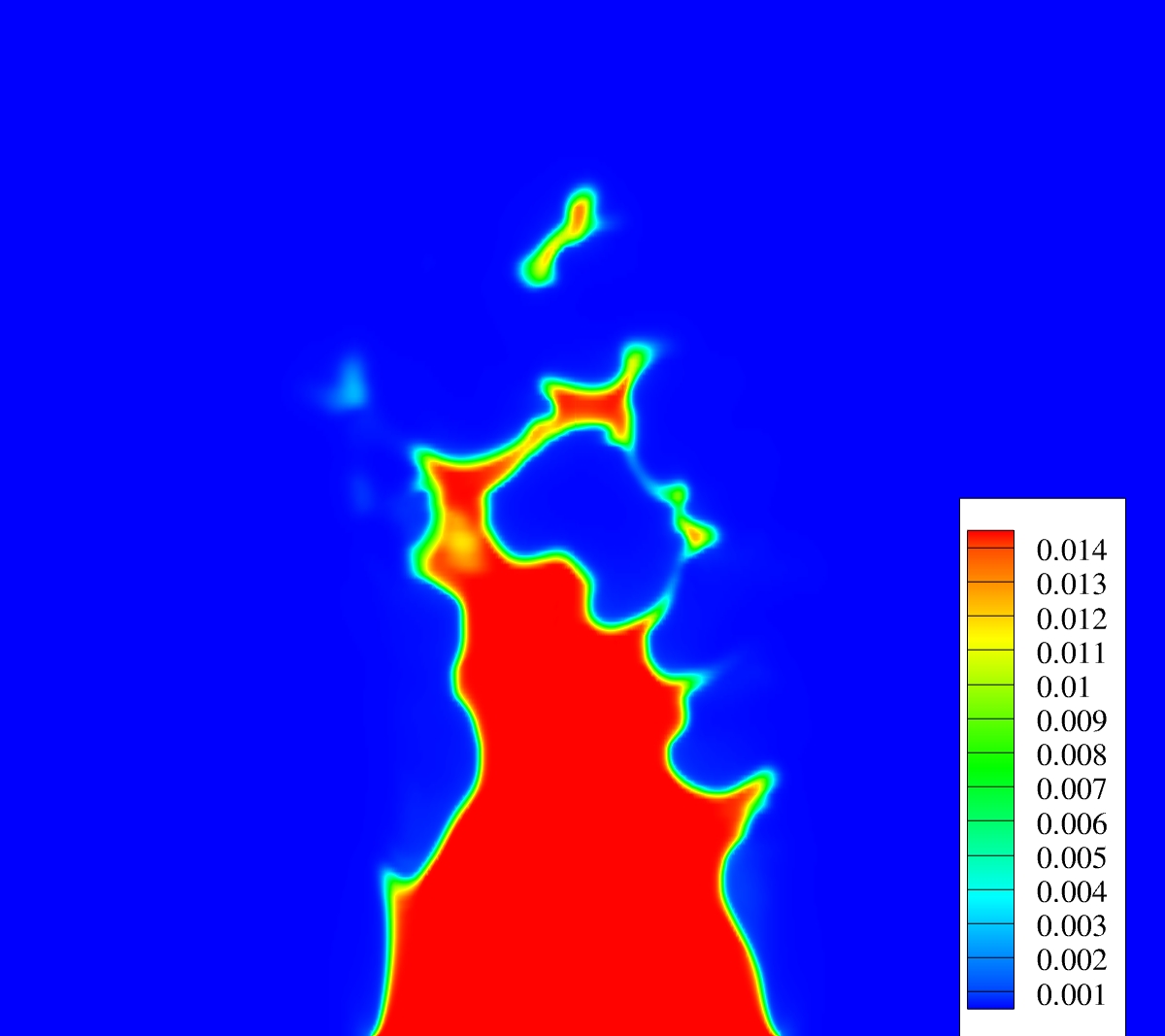}
\\[.05\linewidth]
\includegraphics[width=0.5\linewidth]{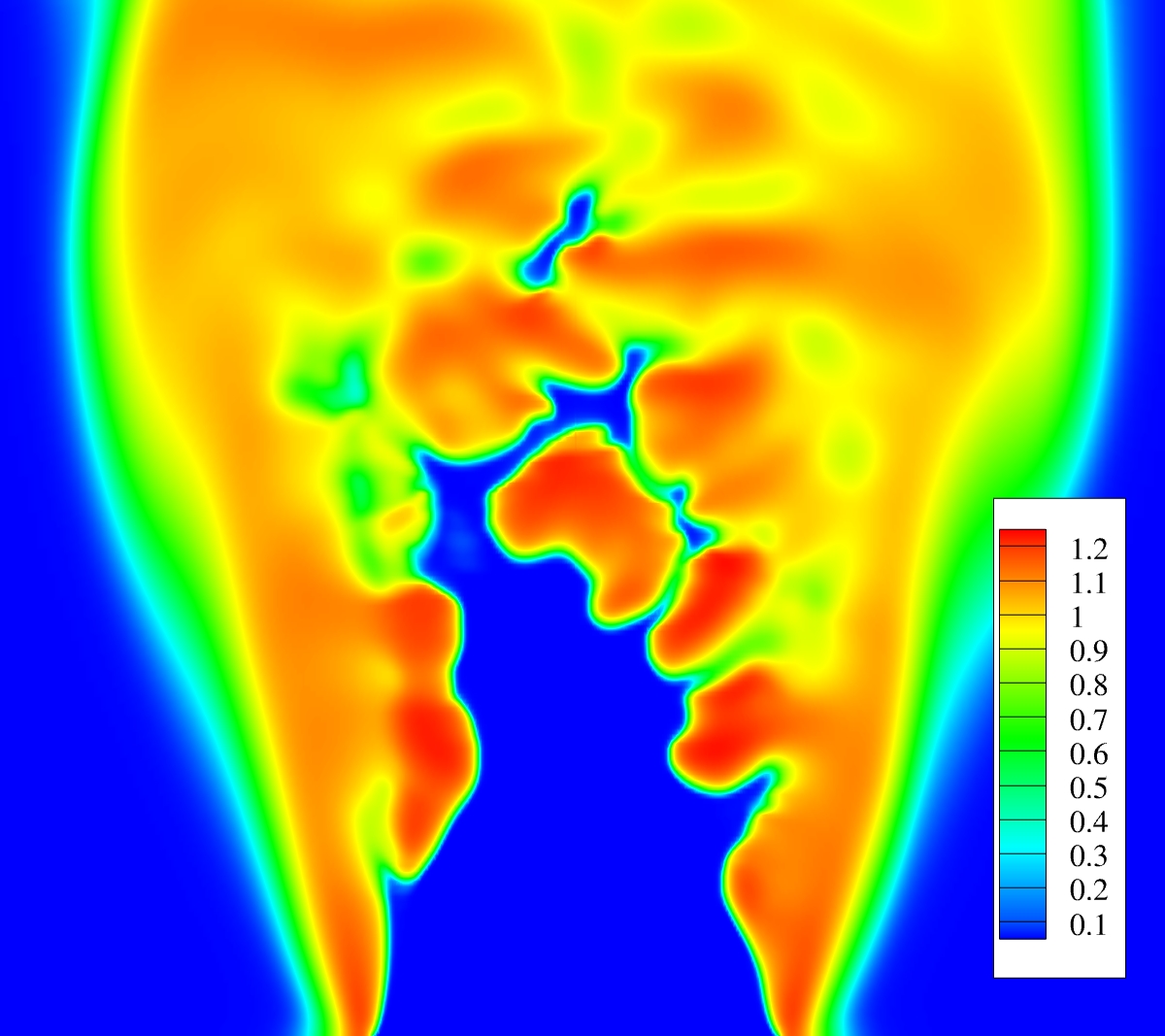}
\caption{\label{fig:1}
2D snapshots of the concentration of molecular hydrogen $Y_{{\rm H}_2}$, top panel, 
and of the progress variable based on the temperature 
$c_T=T-T_u/(T_{ad}-T_u)$ ($T$ temperature, $T_{ad}$ adiabatic 
flame temperature, $T_u$  unburned mixture temperature), bottom 
panel. 
}
\end{figure}
\begin{figure}[t!]
\centering
\includegraphics[width=0.5\linewidth]{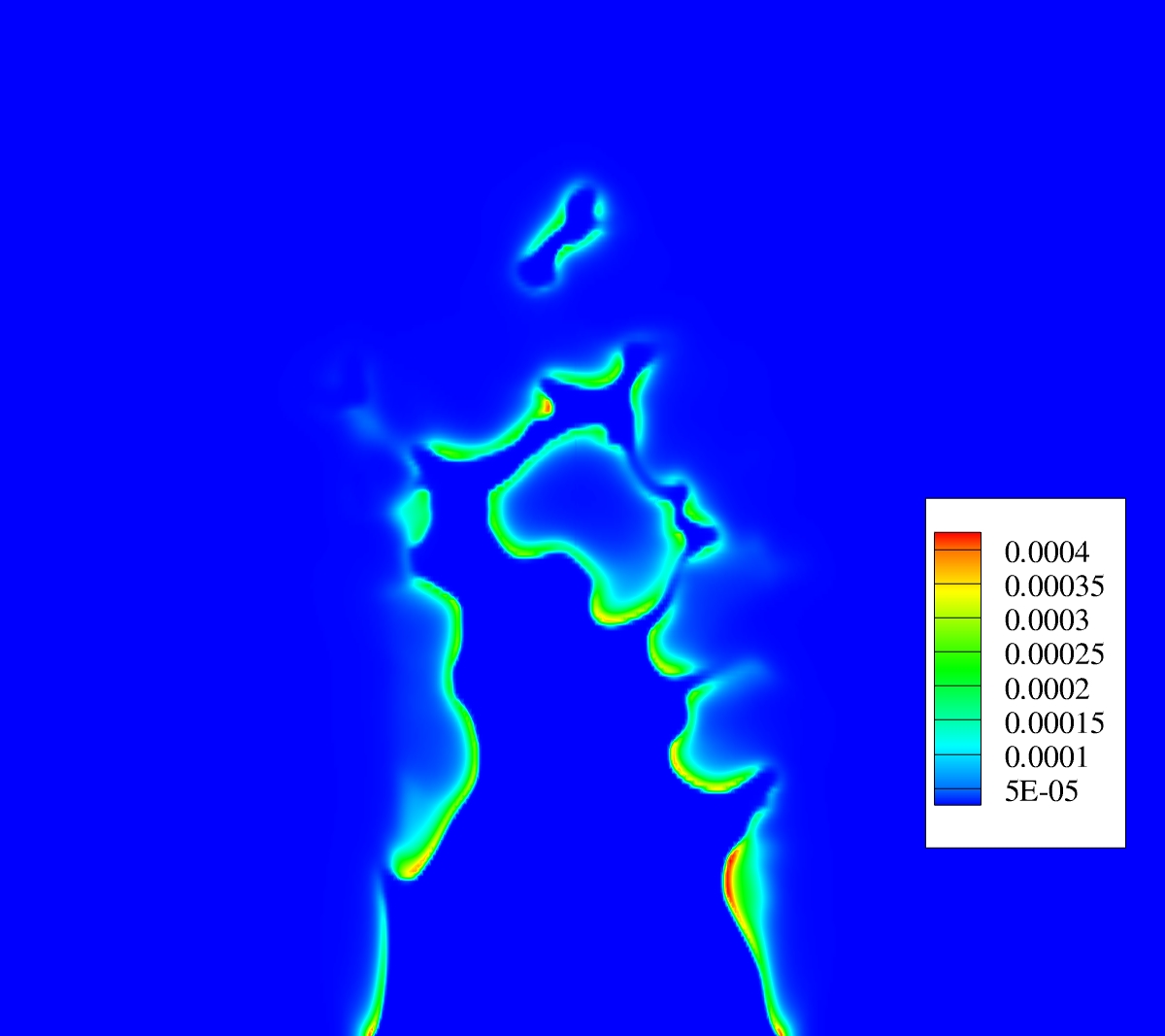}
\\[.05\linewidth]
\includegraphics[width=0.5\linewidth]{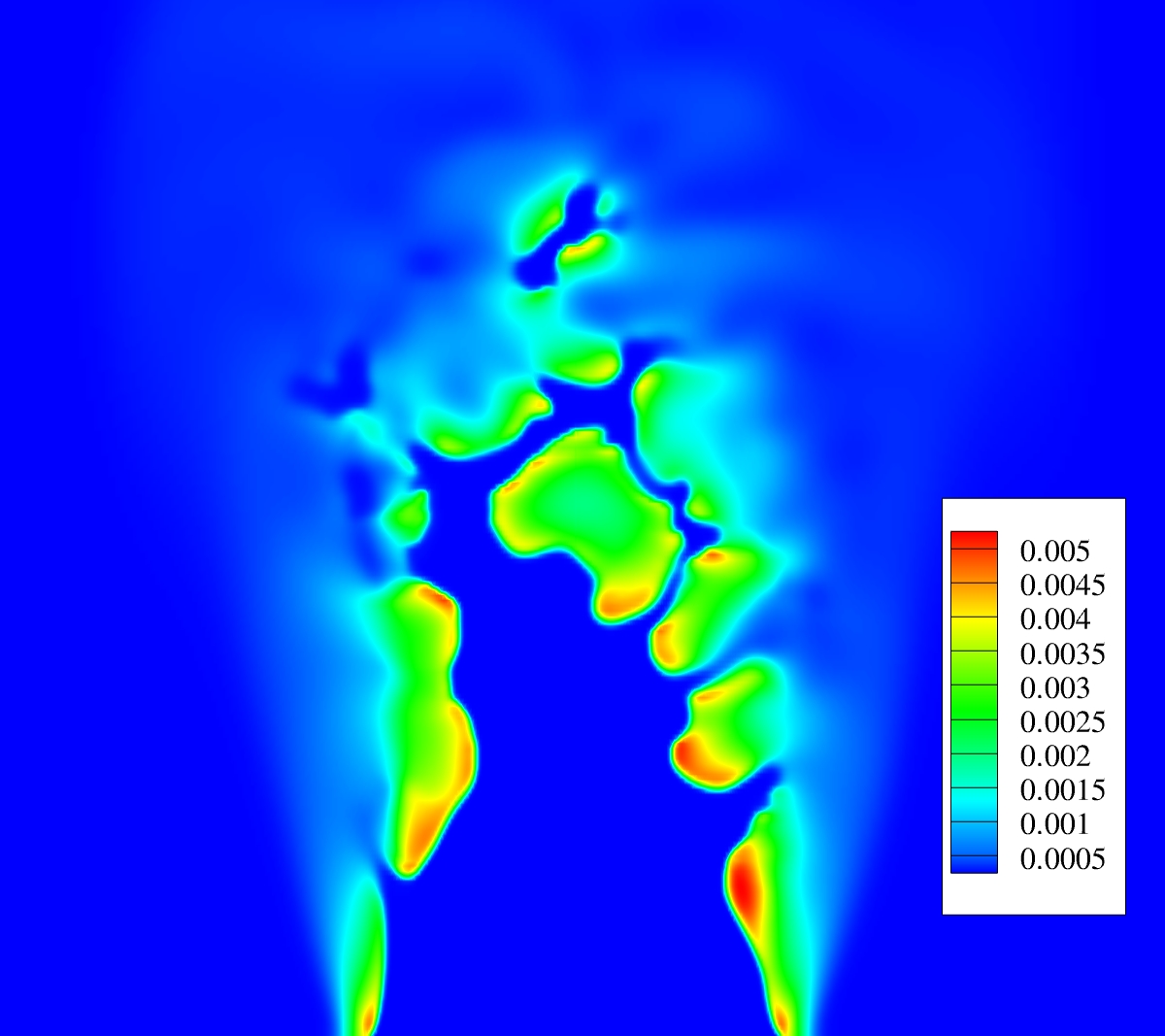}
\caption{\label{fig:2}
2D snapshots of the concentrations of the radicals $Y_{H}$  and $Y_{OH}$, 
top and bottom panels, respectively. 
}
\end{figure}

\subsection{Instantaneous configurations}
\label{ssec:instantaneous}

This section describes the instantaneous behavior of the turbulent flame, focusing 
 on the interaction between turbulence and chemistry 
and the effects of preferential diffusion.

The top panel of Fig.~\ref{fig:1} shows the contours of the 
molecular hydrogen concentration $Y_{{\rm H}_2}$, on a plane containing 
the jet axis extracted from the instantaneous three-dimensional field. The thin green layer 
corresponds to the flame front; $Y_{{\rm H}_2}$ shows 
large variations along the front, moving from the fresh mixture to the products. 
The front is severely wrinkled by the turbulence, to the extent 
that pockets of fresh gas are turned off and carried into the 
product side where they may be eventually consumed. 
The flame front can be parametrized in terms of  a temperature-based 
progress variable $c_T=(T-T_u)/(T_{ad}-T_u)$, where $T_{ad}$
is the adiabatic flame temperature of the inlet mixture and $T_u$ is the 
 temperature of the unburnt mixture.

\begin{figure}[t!]
\centering
\includegraphics[width=0.95\linewidth]{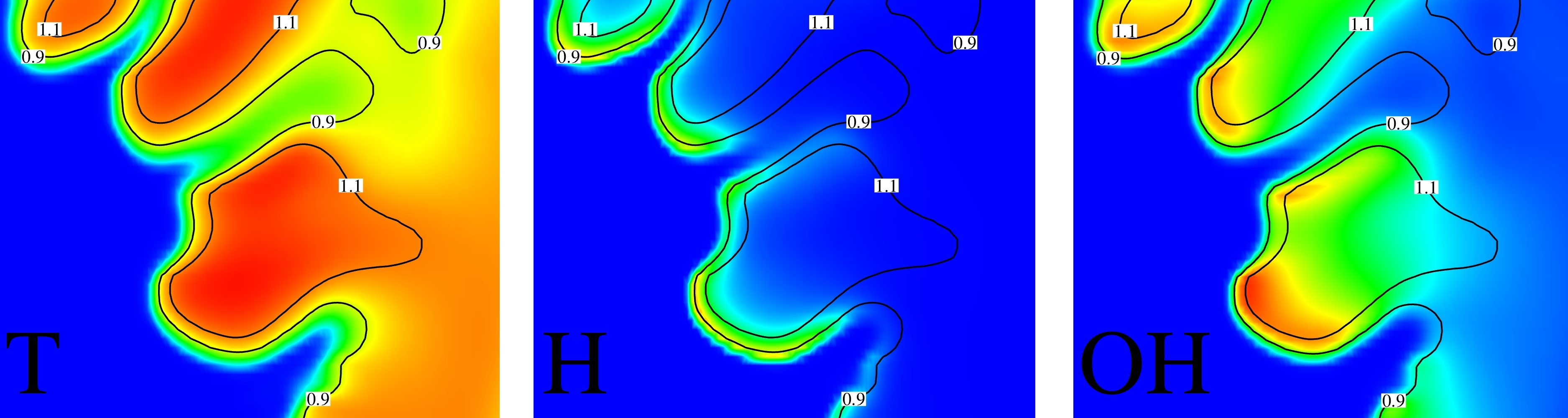}
\caption{\label{fig:3}
2D snapshot enlargements of $T$ (left),  $Y_{\rm H}$
(middle)  and $Y_{\rm OH}$ (right). Isolines are two isolevels of $c_T$: 0.9 and 1.1, 
respectively. The color maps match those of figure~\ref{fig:1} and
\ref{fig:2}.
}
\end{figure}

Two different types of structures identify the flame front:
the {\sl gullies}, i.e. the 
narrow regions concave towards  the fresh gas, and the {\sl bulges}, 
i.e. the larger, smoother regions convex towards the fresh 
gas~\cite{chebil1,chebil2}.
The bulges are characterized by values of $c_T$ significantly exceeding unity, 
 up to  1.2 (super-adiabaticity), whereas sub-adiabatic conditions, 
$c_T<1$, occur in gullies, even in regions where the hydrogen concentration 
$Y_{{\rm H}_2}$ is very low. In both regions,  the  isolevels of   $Y_{\rm H}$, $Y_{\rm OH}$
follow the flame front closely, as shown in  figure~\ref{fig:2}, confirming their role as  
good markers of the progress of combustion, see e.g. ~\cite{chebil1,frakalbil,stabillyofralon}. 
The  $Y_{\rm H}$ concentration profile appears to be symmetric 
with respect to the peak concentration values and tends to be
localized near the flame front.  $Y_{\rm OH}$ displays 
a completely different behavior since it sharply 
grows at  the beginning of the flame front and slowly decreases towards the hot gas region,
so that OH radical
 persists well inside the burnt region.
The high concentration of radicals in the bulges attests to the intense chemical activity in these regions,
while their absence in the gullies is an indication of flame quenching. 
Figure~\ref{fig:3} shows  two isolines of the 
progress variable $c_T$ overlapping with the contour plot of the concentrations of the radicals. 
The almost exact parallelism of the isolines resembles  the lamellar structure observed in the experimental results of 
Chen and Bilger~\cite{chebil1,chebil2}. The isolines appear to be closer 
in the bulges than in the gullies, indicating  steeper temperature 
gradients $|\nabla c_T|$, which are related to a reduction 
of the flame thickness $\delta \sim 1/|\nabla c_T|$ and an intensified
scalar dissipation rate $\chi \sim |\nabla c_T|^2$~\cite{chakleswa}. 
Smoother temperature gradients in the gullies are associated with 
reduced concentrations of the radicals and the presence of 
significant quenching effects.
Turbulent fluctuations enhance flame corrugation which, however, is also known to occur in
two dimensional simulations and experiments~\cite{altfrotomkerbou,kadsahastat}  on laminar flames.
Indeed lean hydrogen flames are strongly subject to 
differential diffusion
 and thermo-diffusive instabilities~\cite{barzelist}, which are responsible of the flame wrinkling, 
temperature oscillations and possible local extinctions, independently of the turbulent 
or laminar regime. This type of instability is associated with the high 
diffusivity of the molecular hydrogen H$_2$ and  a consequent very low 
Lewis number, ${\rm Le} =D_T /D_{{\rm H}_2}\ll1$. 
Hydrogen diffuses towards the burnt gases much faster than heat diffuses towards 
the fresh gases. The ratio of hydrogen to temperature diffusion lengths is
the inverse square root of the Lewis number, 
$\delta_{{\rm H}_2}/\delta_T = 1/\sqrt{Le}$. This effect, combined with the local 
curvature of the front, implies the local enrichment of the mixture in the tiny 
diffusively heated region (i.e. the bulges). Consequently, the burnt temperature of the gas 
exceeds the  corresponding temperature of the laminar planar flame.  
In gullies, due to the opposite curvature, reactants diffuse in a 
large zone, resulting in a 
decrease in the temperature of the flame. The presence of both positive and negative 
curvatures induces an unstable situation~\cite{poivey}, which is further enhanced by the turbulent 
fluctuations. 

\subsection{Statistical analysis}

\begin{figure}[h!]
\includegraphics[width=.5\textwidth]{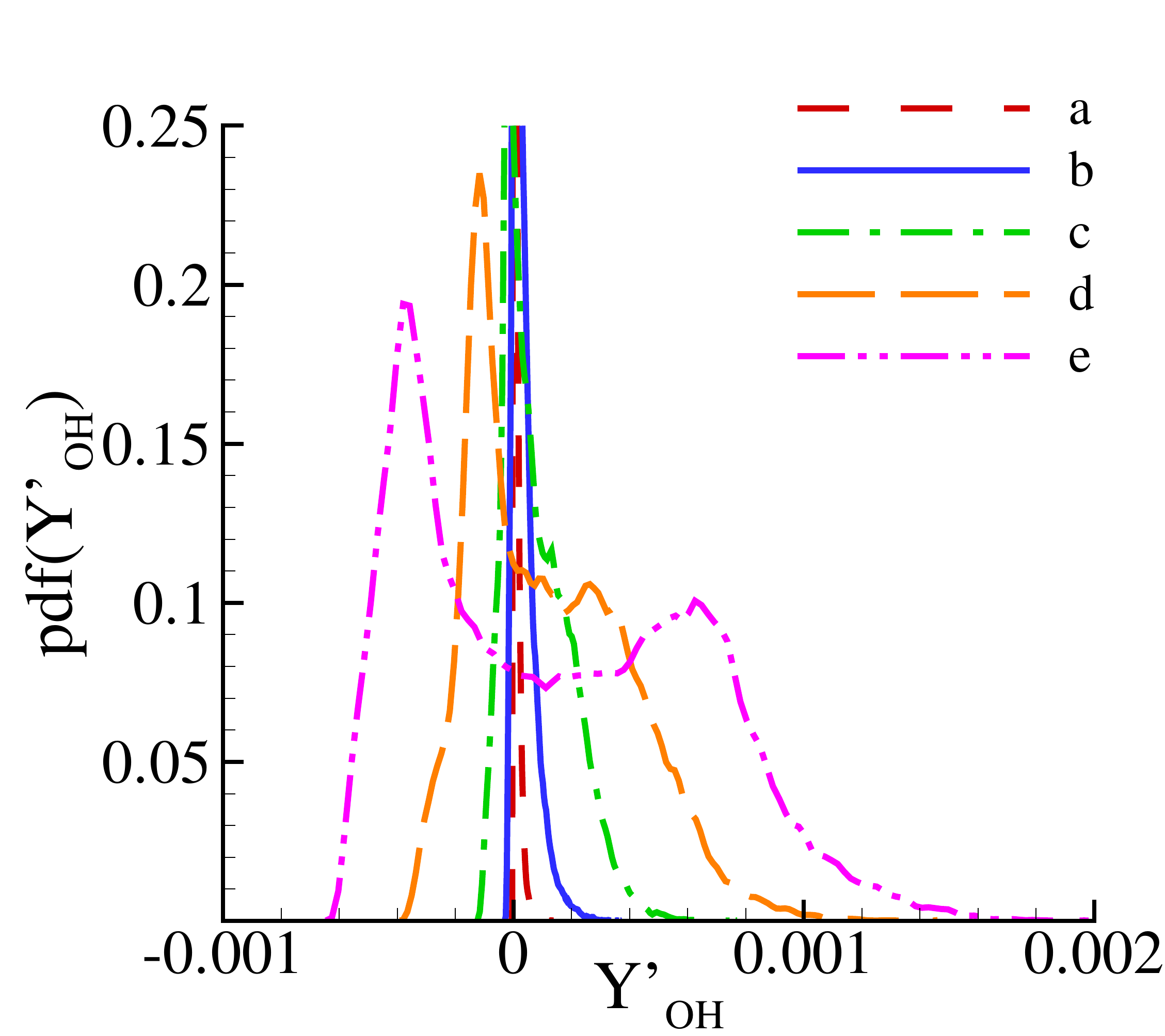}
\includegraphics[width=.5\textwidth]{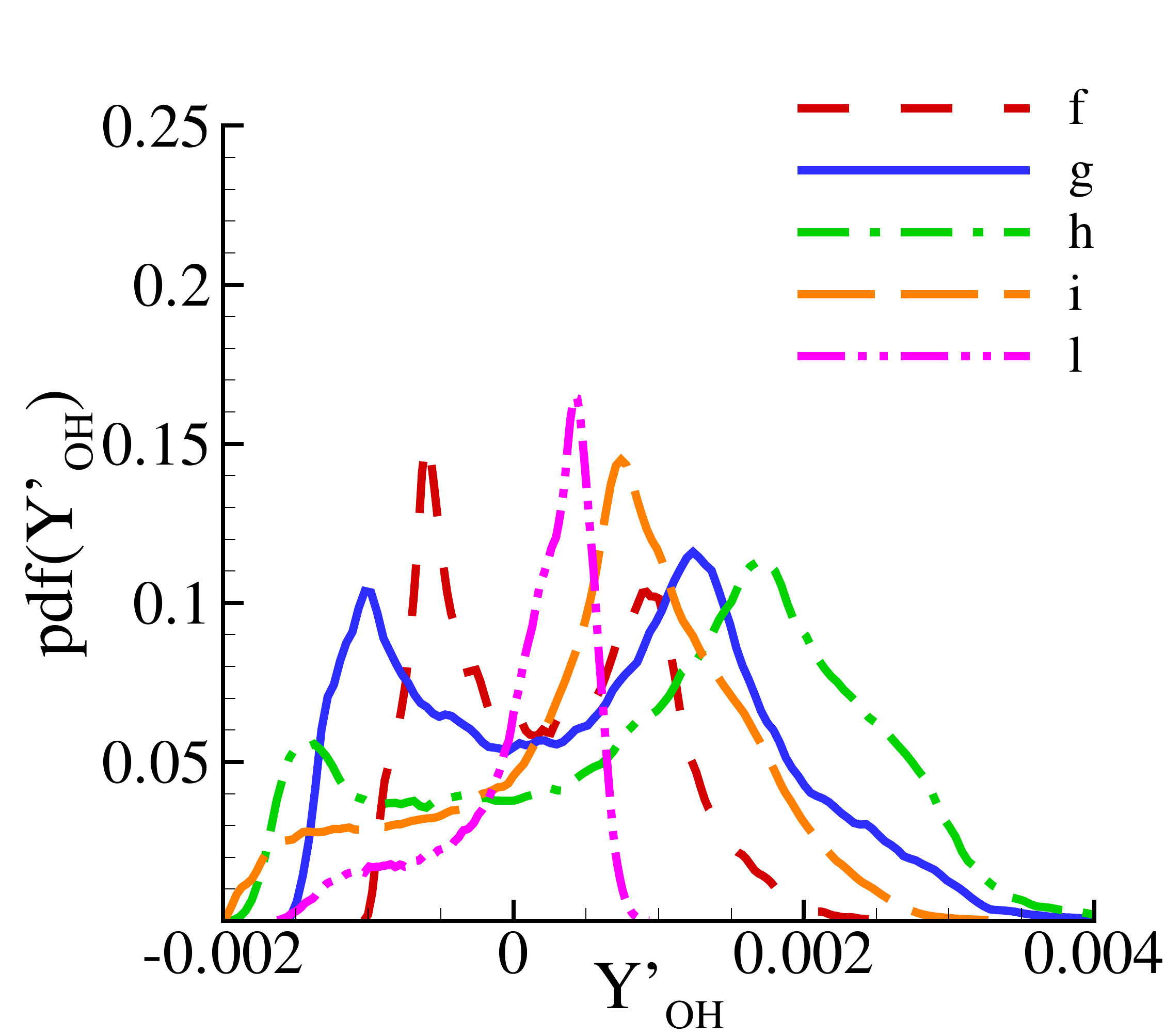}
\caption{\label{fig:5a} 
Probability density function of {the OH radical mass fraction flamelet fluctuation,
see eq.~\eqref{e:y'_oh}}, at different H$_2$ concentration intervals: 
(a) $1\times10^{-2} \le Y_{\rm{H}_2} \le1.2\times10^{-2}$.
(b) $8\times10^{-3} \le Y_{\rm{H}_2} \le  1\times10^{-2}$;
(c) $6\times10^{-3} \le Y_{\rm{H}_2} \le  8\times10^{-3}$;
(d) $4\times10^{-3} \le Y_{\rm{H}_2} \le  6\times10^{-3}$;
(e) $3\times10^{-3} \le Y_{\rm{H}_2} \le  4\times10^{-3}$;
(f) $2\times10^{-3} \le Y_{\rm{H}_2} \le  3\times10^{-3}$;
(g) $1\times10^{-3} \le Y_{\rm{H}_2} \le  2\times10^{-3}$;
(h) $5\times10^{-4} \le Y_{\rm{H}_2} \le  1\times10^{-3}$;
(i) $1\times10^{-4} \le Y_{\rm{H}_2} \le  5\times10^{-4}$;  
(l) $5\times10^{-5} \le Y_{\rm{H}_2} \le  1\times10^{-4}$.
}
\end{figure}

The instantaneous fields showed that combustion in lean hydrogen flames 
is mainly characterized by two different states: more reactive regions, exhibiting super-adiabatic
temperatures (bulges), alternated with nearly quenched areas (gullies), which display lower temperatures.
A global characterization, which shows
the statistical weight of these two states, can be given in terms of the probability
density function (pdf) of the flamelet fluctuation of OH concentration, $Y'_{\rm OH}$, eq.~\eqref{e:y'_oh}, as shown in 
figure ~\ref{fig:5a}.

As explained in section \S~\ref{sec:nummet}, $Y'_{\rm OH}$ is 
defined as the difference of local turbulent concentration of radical OH with that corresponding 
the same $Y_{{\rm H}_2}$ in the unstretched laminar flame. 

Figure~\ref{fig:5a} shows the probability density function (pdf) of $Y'_{\rm OH}$ 
conditioned to different intervals of ${\rm H}_2$ concentration. This procedure allows to analyze different regions of the
instantaneous reaction zone moving from the fresh gas
 (high $Y_{\rm H_2}$) to the burnt region (low $Y_{\rm H_2}$).
Close to the fresh gas region, plots (a), (b), (c), and (d) in the left panel of figure~\ref{fig:5a}, the pdf is 
characterized by a mono-modal behavior with a slightly negative most frequent state and an intense 
positive tail. In this part of the flame front the combustion process is reasonably 
described by a flamelet approximation 
with the local flame structure similar to the unstretched laminar flame, note the pdf close to a Dirac 
function centered at the origin. 
Despite the similarity, the steep positive tails denote intense intermittent events with strong 
chemical activity
that cannot be described in the standard flamelet framework.  
Proceeding across the flame, plots (e) of left panel and (f), (g), 
and (h) of right panel, the pdf of $Y'_{\rm OH}$ shows a bi-modal distribution with one 
maximum in correspondence with negative and the other with positive fluctuations, respectively.
This distribution is inconsistent with the pure flamelet regime whose pdf should appear as a 
narrow mono-modal distribution centered at the origin.
Indeed the flamelet regime corresponds to turbulent flames wrinkled by turbulent 
fluctuations that locally share the structure of a laminar flame. The present results provide 
evidence that the flame structure is typically inconsistent with a flamelet.
Focusing on plot (g) of fig.~\ref{fig:5a} for definitness, 
 the two modal values of the distribution correspond to slow 
($Y'_{\rm OH}\simeq-0.001$) and fast ($Y'_{\rm OH}\simeq0.0012$) burning states. They 
occur almost $50\%$ and $90\%$ more frequently than  the unstrained laminar state 
($Y'_{\rm OH}=0$). 
The state $Y'_{\rm OH}\simeq-0.001$ appears to be associated with almost quenched areas since the 
local radical concentration $Y_{\rm OH} \simeq  Y'_{\rm OH} + \bar{Y}^L_{\rm OH}$ 
almost vanishes ($\bar{Y}^L_{\rm OH}\sim 0.001$ when 
$1\times10^{-3} \le Y_{\rm{H}_2} \le  2\times10^{-3}$). 
In the fast burning state, the concentration of OH turns out to be more than twice the laminar value.
On the contrary, in the last stage of the turbulent reaction zone (towards the burnt gases), 
plots (i) and (l) in the right panel of Fig.~\ref{fig:5a}, 
a mono-modal behavior is recovered, though characterized by a positive mode and an 
intense negative tail. Indeed, this reaction region is seemingly characterized by 
higher chemical activity than the laminar flamelet and by strong intermittent events.

The evidence is that the current turbulent lean hydrogen flame can be 
sketched as a bi-stable system with two distinct states reacting more slowly (nearly quenched)
and more quickly than the laminar unstretched flamelet. 
These events are prevailing in the inner part of the flame and influence the fore and hind 
reaction region by intermittent intense-burning and quenching events, respectively (pdf tails).

In order to discuss the strong impact that turbulent fluctuations have on the 
dynamics of the flame, we address the correlations among 
flamelet fluctuations of the radical $OH$, velocity gradient and flame curvature. 
\begin{figure}[t!]
\centering
\includegraphics[width=.32\textwidth]{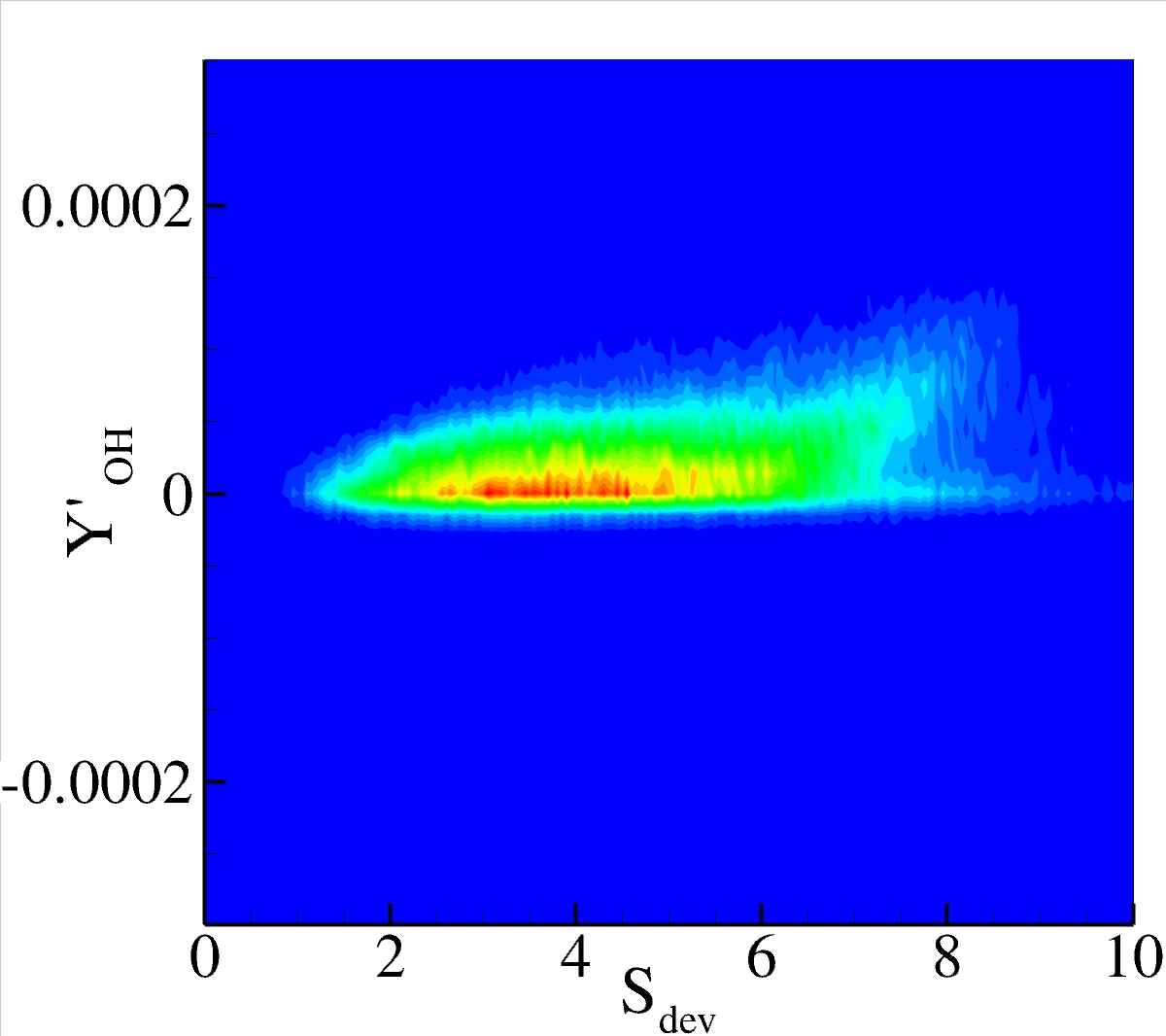}
\includegraphics[width=.32\textwidth]{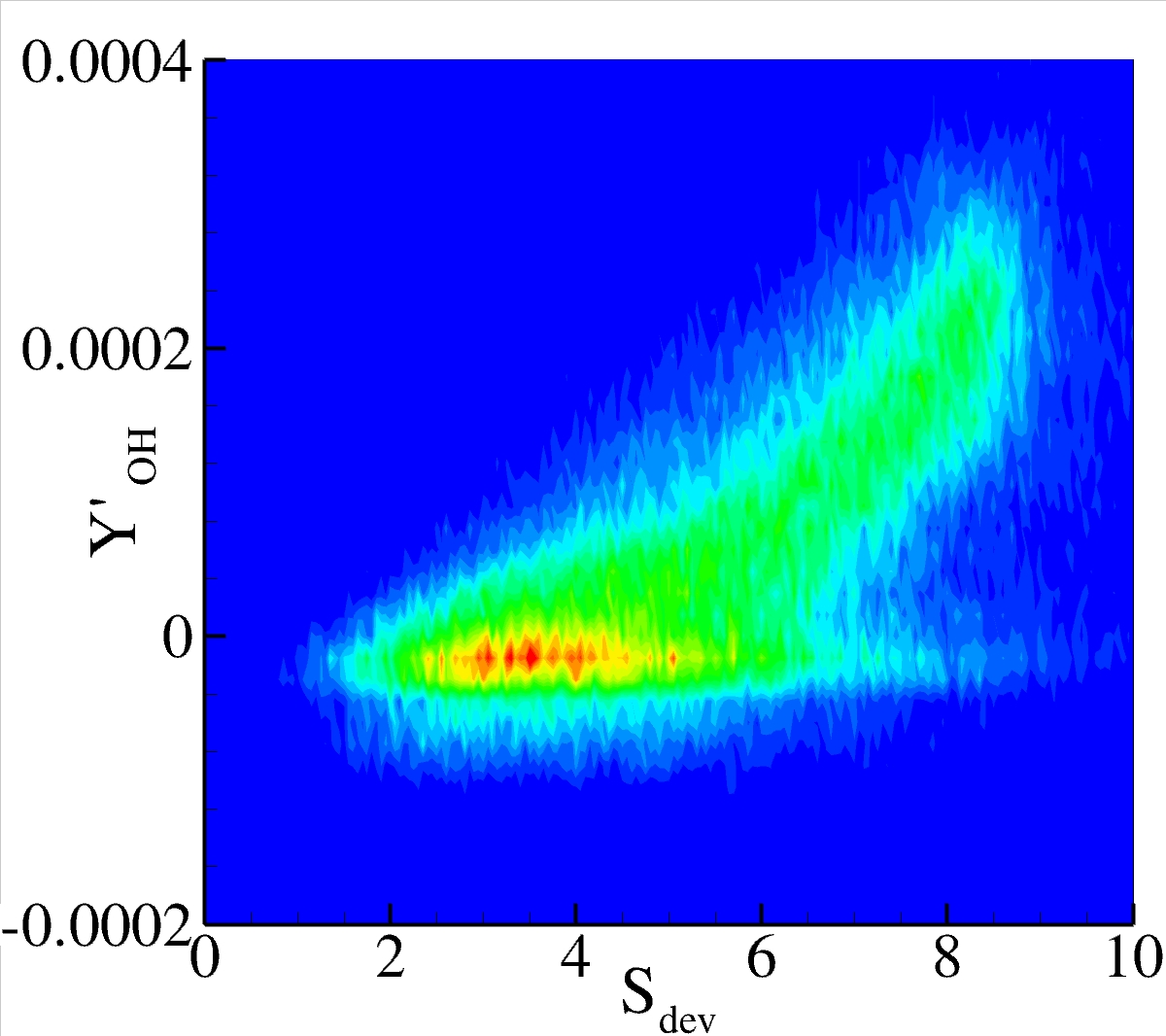}
\includegraphics[width=.32\textwidth]{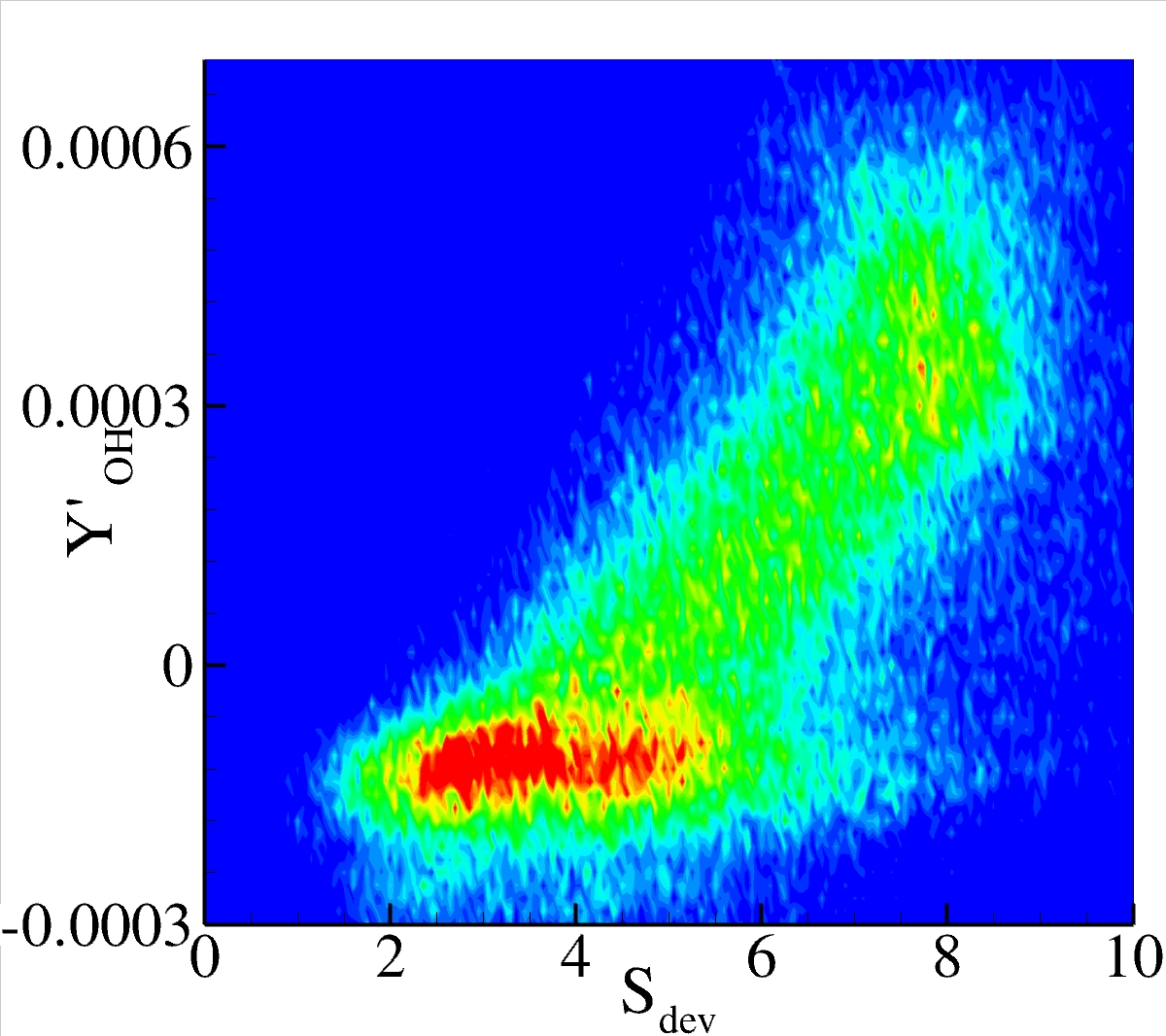}
{\scriptsize \put(-330,100){\bf (b)}}
{\scriptsize \put(-215,100){\bf (c)}}
{\scriptsize \put(-100,100){\bf (d)}}\\
\includegraphics[width=.32\textwidth]{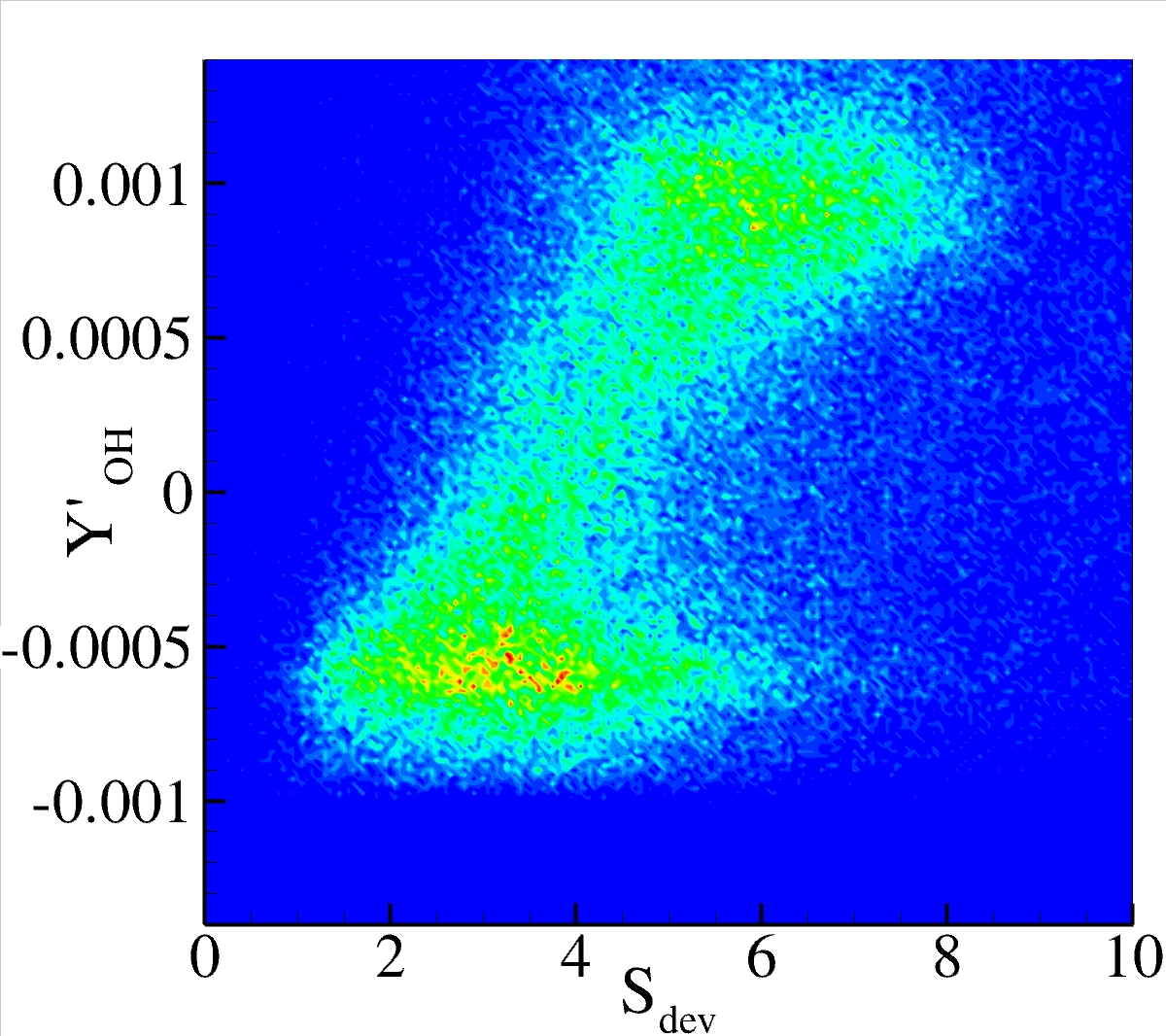}
\includegraphics[width=.32\textwidth]{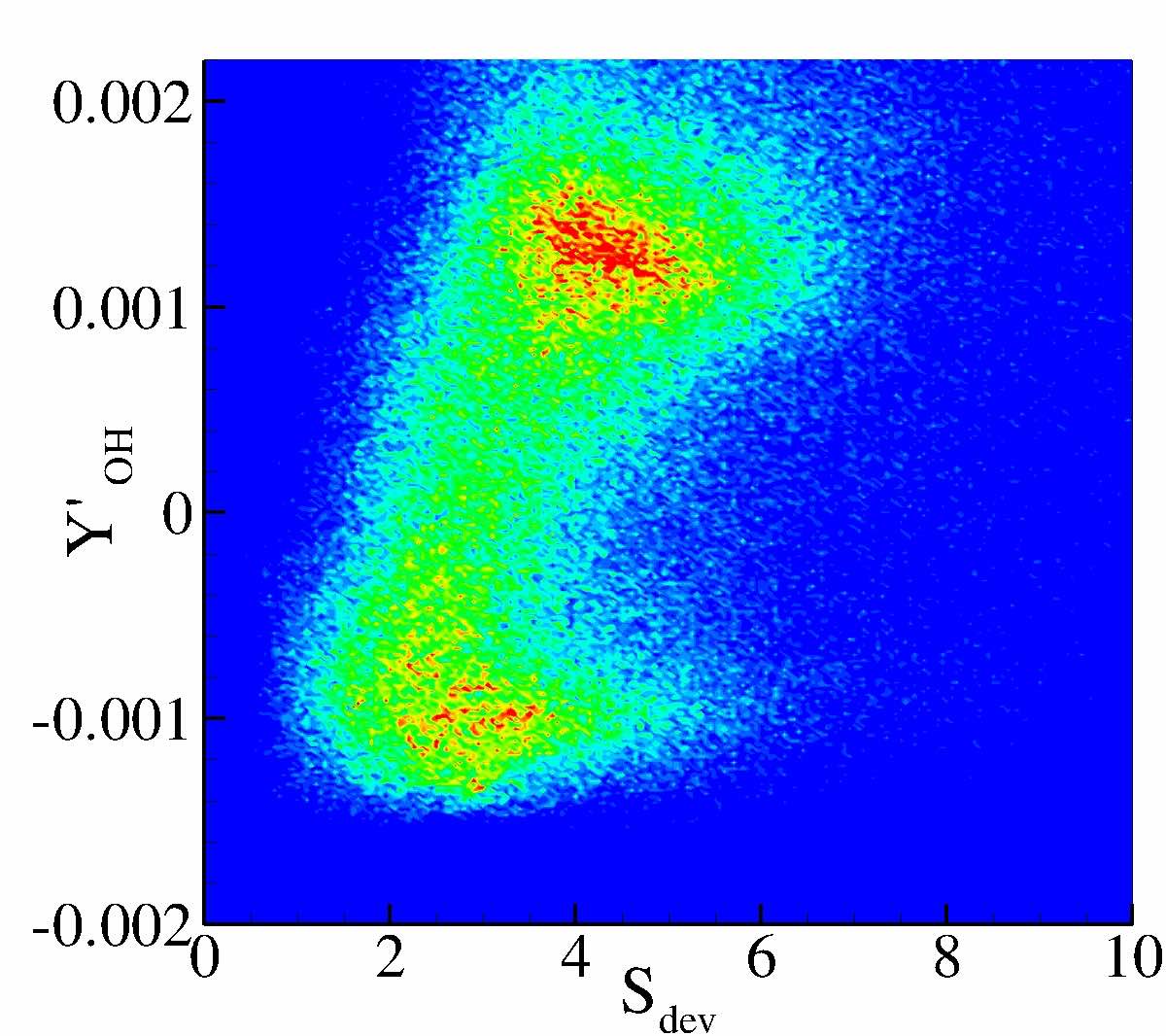}
\includegraphics[width=.32\textwidth]{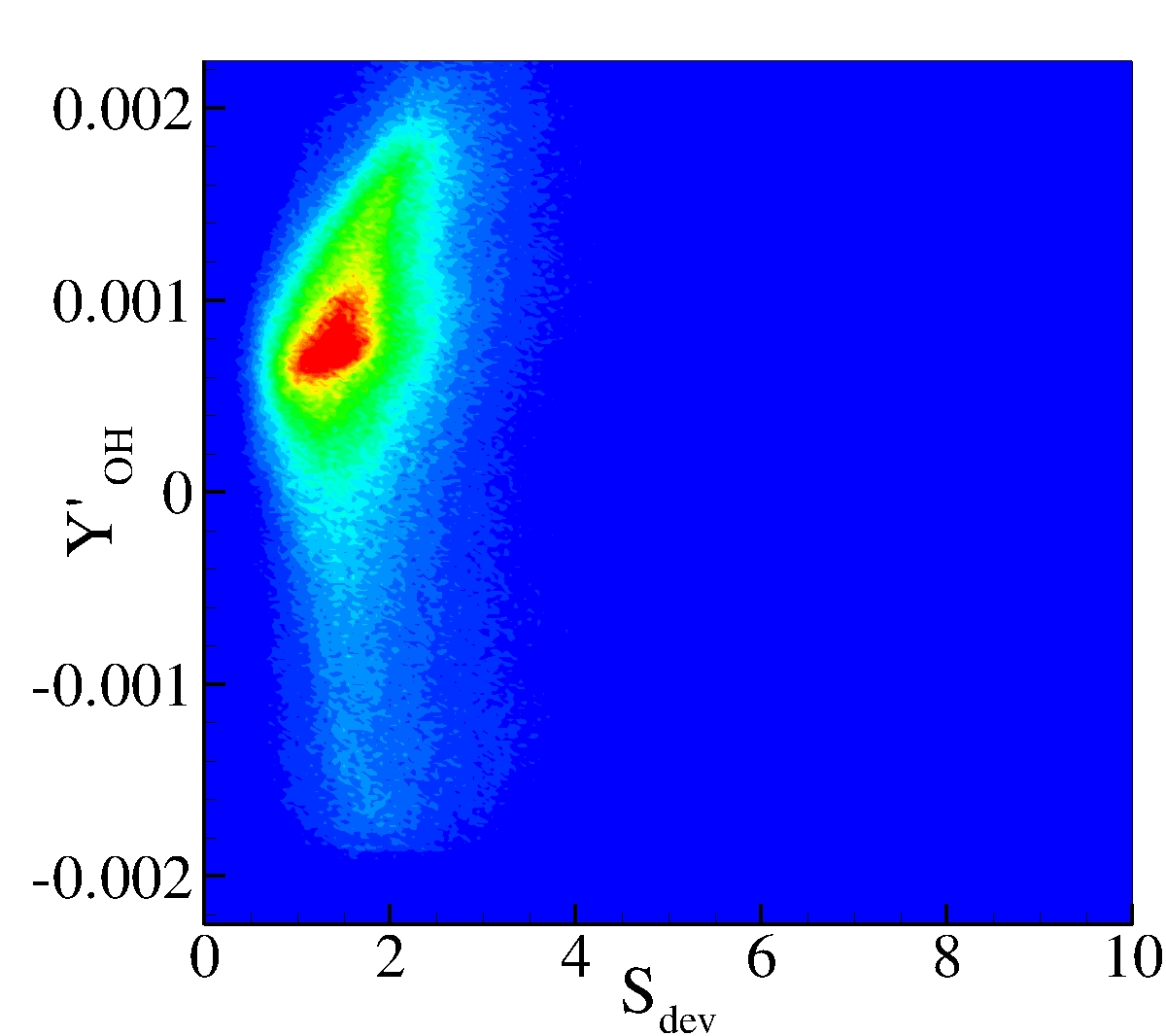}
{\scriptsize \put(-330,100){\bf (f)}}
{\scriptsize \put(-215,100){\bf (g)}}
{\scriptsize \put(-100,100){\bf (i)}}\\
\caption{\label{fig:5} Joint probability density function (jpdf) of
the flamelet normalized OH concentration $Y'_{\rm OH}$ and the deviatoric 
strain rate $S_{dev}=\sqrt{E_{dev}:E_{dev}}$, normalized by $R/U_0$. 
In order to avoid spurious sampling effects, jpdf are evaluated along 
the flame front conditioning at different H$_2$ concentration intervals: 
(b) $8\times10^{-3} \le Y_{\rm{H}_2} \le  1\times10^{-2}$;
(c) $6\times10^{-3} \le Y_{\rm{H}_2} \le  8\times10^{-3}$;
(d) $4\times10^{-3} \le Y_{\rm{H}_2} \le  6\times10^{-3}$;
(f) $2\times10^{-3} \le Y_{\rm{H}_2} \le  3\times10^{-3}$;
(g) $1\times10^{-3} \le Y_{\rm{H}_2} \le  2\times10^{-3}$;
(i) $1\times10^{-4} \le Y_{\rm{H}_2} \le  5\times10^{-4}$.
 }
\end{figure}
Given the strong gas expansion in the flame, the effect of local 
deformation induced by turbulence could be blurred by the flow divergence. 
In these conditions, the statistical characterization of the turbulence-combustion interaction
interaction is better achieved by considering the deviatoric part of the strain rate  
$\mathbf{E_{dev}}=\left(\nabla \mathbf{u}+\nabla \mathbf{u}^T \right)/2-\left(\nabla \cdot \mathbf{u}\right)\mathbf{I}/3$, 
as it appears in the equation for the local stretching rate $\varkappa$~\cite{poivey},
\begin{equation}\label{stretch}
\varkappa=\frac{1}{A}\frac{dA}{dt}=\underbrace{\frac{2}{3}\nabla  \cdot \mathbf{u}}_{I}-\underbrace{\mathbf{n}\otimes \mathbf{n}:\mathbf{E_{dev}}}_{II}+\underbrace{S_d \, (\nabla \cdot \mathbf{n})}_{III} \ .
\end{equation}
The stretching rate incorporates three different kinematic effects. Besides the gas expansion, which is 
directly associated with heat released during the combustion (term $I$), two additional terms on the 
right hand side of equation (\ref{stretch}) contribute to the stretching rate: the strain effect,
(term $II$), and the mean curvature of the reaction  front ($III$). 
In accordance with equation (\ref{stretch}), we have found that 
these last two terms mostly influence the flame dynamics and they both clearly demonstrate 
correlation with the chemical activity. 

Figure \ref{fig:5} shows the joint-pdf of deviatoric strain rate magnitude, 
$\mathcal{S}_{dev}=\sqrt{2 \mathbf{E_{dev}}:\mathbf{E_{dev}} }$, and flamelet 
fluctuation of the $OH$ radical concentration, $Y'_{\rm OH}$. 
{Statistics are presented for six of the ten conditional $H_2$ concentrations used 
in figure~\ref{fig:5a}}, in particular, the intervals of $Y_{\rm{H}_2}$ denoted by the letters (b), (c), (d), 
(f), (g), and (i).
The aft part of the flame (towards the fresh mixture, high H$_2$ concentration) is 
substantially unaffected by the deviatoric strain, with OH radical concentration following the 
behavior of the unstretched, laminar flamelet. 
Moving inside the reaction region ((c) and (d) panels) an high deviatoric strain promotes intense 
chemical activity, i.e. $Y'_{\rm OH}$ is positive.
In the hind part of the flame (panels (f) and (g)), the positive $Y'_{\rm OH} - S_{dev}$ correlation 
indicates that strong burning states are associated with high deviatoric strain rates.
This correlation reduces moving further towards the burnt region to become very weak in 
panel (i) where the mixture is almost entirely burnt.
The positive mode of $Y'_{\rm OH}$ clearly shows the higher OH radical concentration found in the
outer part of the turbulent flame (product side) in comparison with the reference flamelet.

\begin{figure}[t!]
\centering
\includegraphics[width=.32\textwidth]{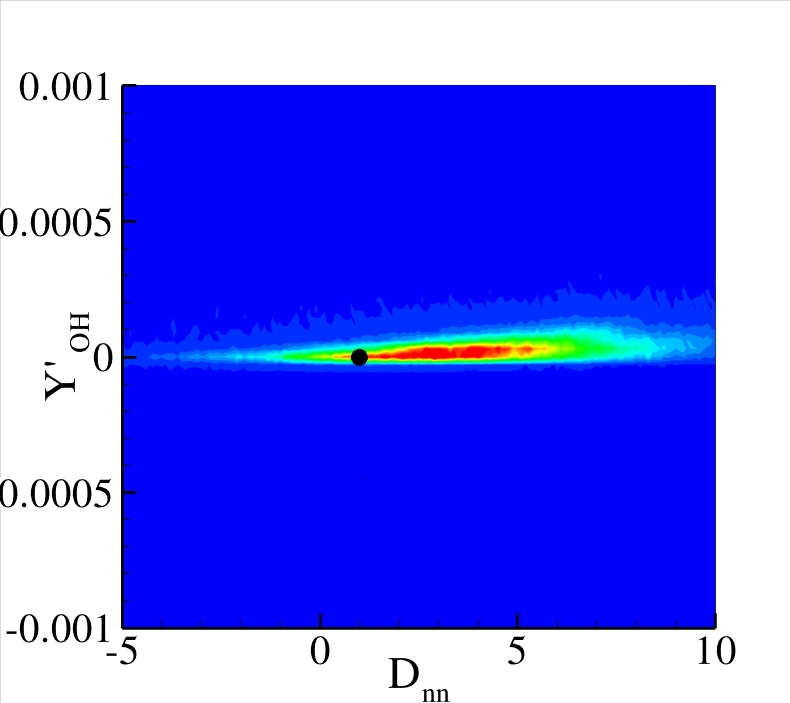}
\includegraphics[width=.32\textwidth]{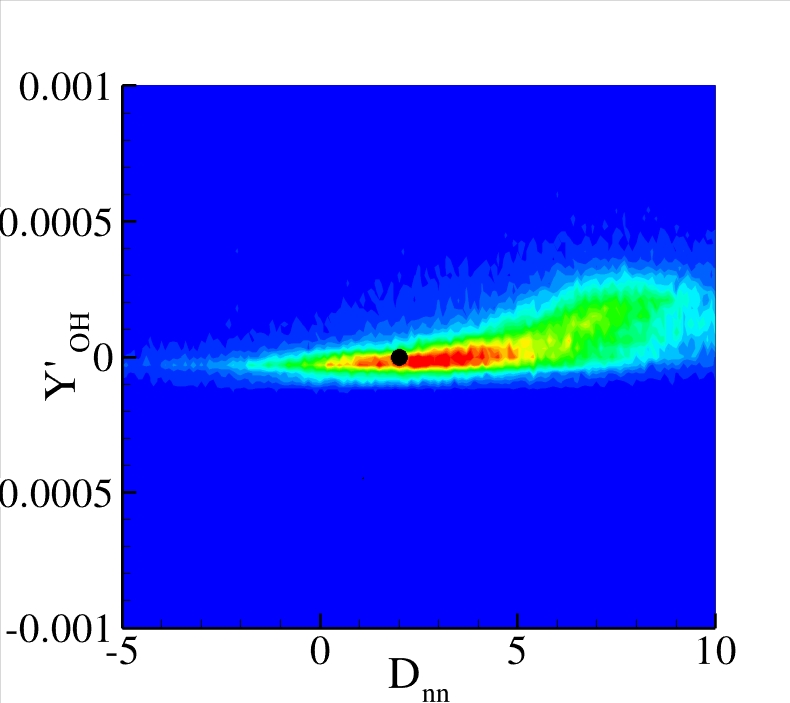}
\includegraphics[width=.32\textwidth]{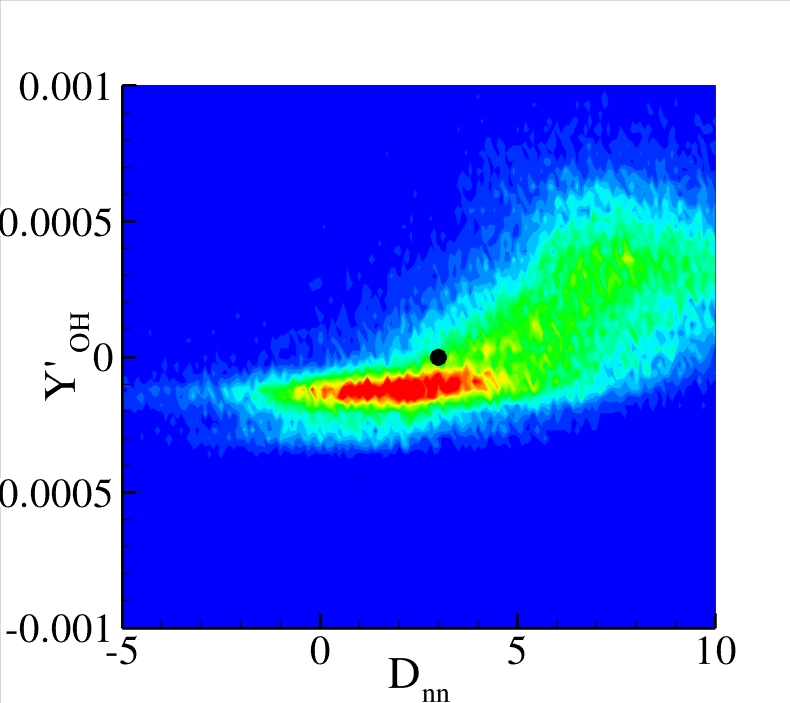}
{\scriptsize \put(-330,100){\bf (b)}}
{\scriptsize \put(-215,100){\bf (c)}}
{\scriptsize \put(-100,100){\bf (d)}}\\
\includegraphics[width=.32\textwidth]{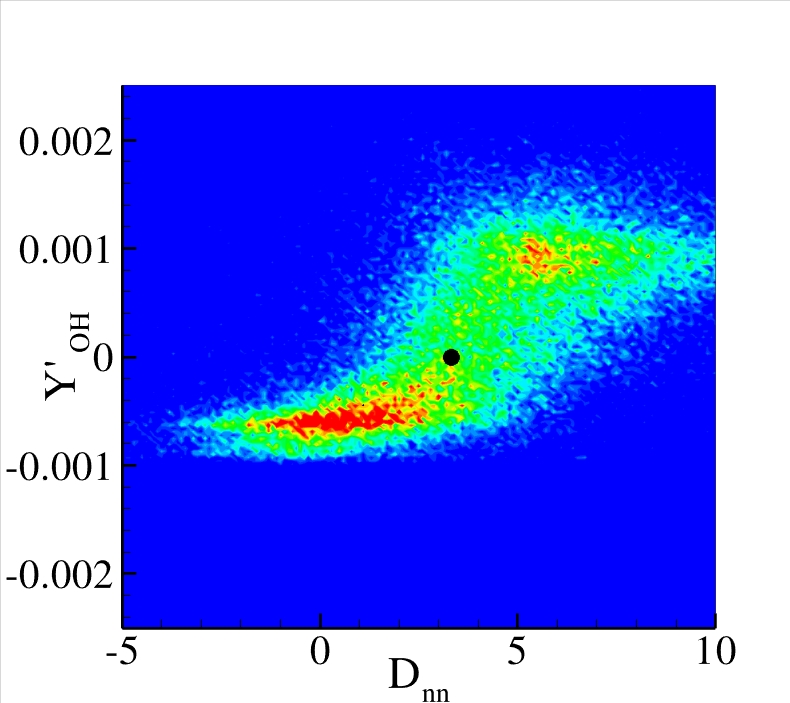}
\includegraphics[width=.32\textwidth]{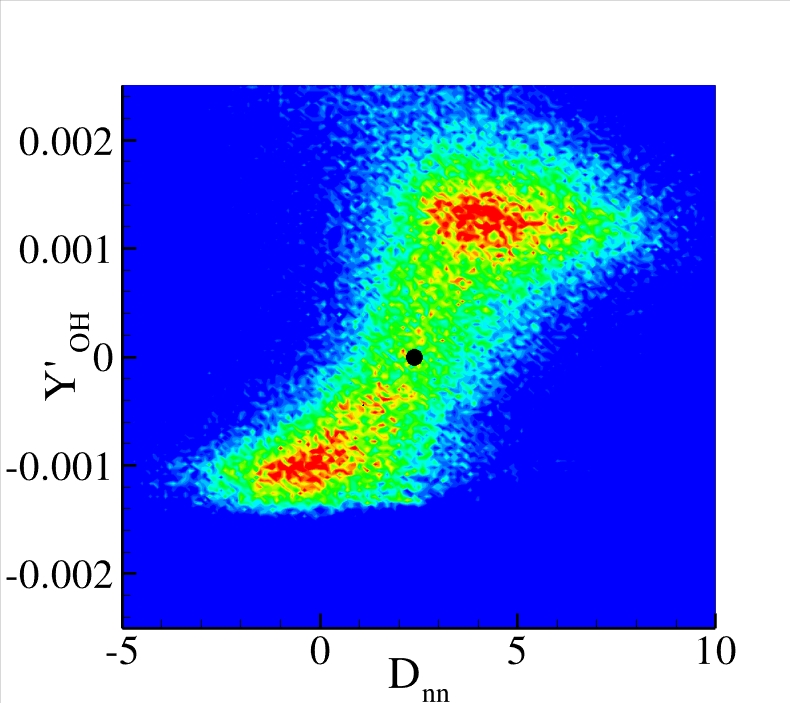}
\includegraphics[width=.32\textwidth]{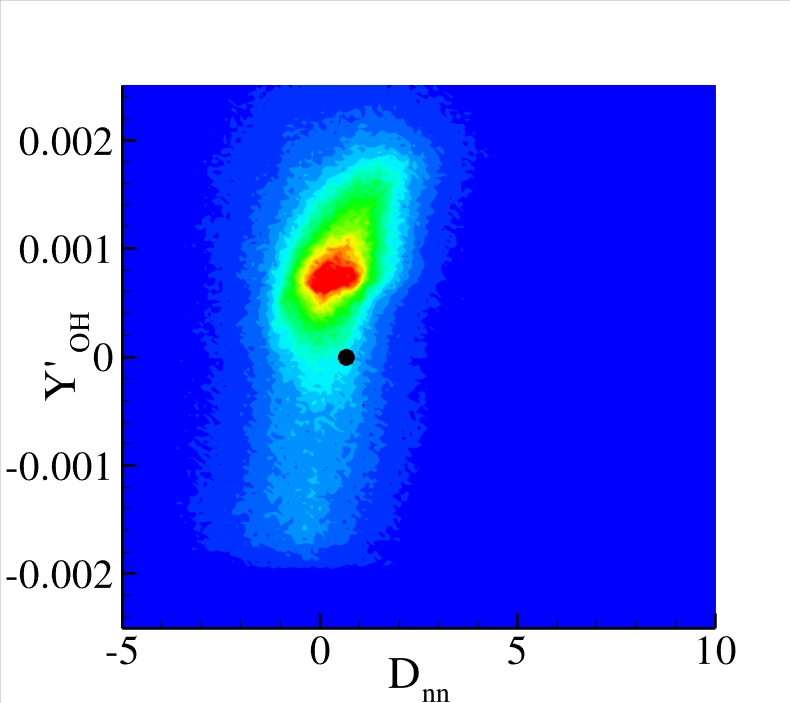}
{\scriptsize \put(-330,100){\bf (f)}}
{\scriptsize \put(-215,100){\bf (g)}}
{\scriptsize \put(-100,100){\bf (i)}}\\
\caption{\label{fig:6}
Joint probability density function (jpdf) in the flame front 
of the flamelet normalised OH concentration $Y'_{\rm OH}$ 
(see eq.~\eqref{e:y'_oh}) and the stress normal to the flame front 
(deviatoric part) $D_{n\,n}=E_{dev}:\bf n\, n$,  
The black circle represents the unstretched laminar flame.
Different intervals of H$_2$ concentration are considered: 
(b) $8\times10^{-3} \le Y_{\rm{H}_2} \le  1\times10^{-2}$;
(c) $6\times10^{-3} \le Y_{\rm{H}_2} \le  8\times10^{-3}$;
(d) $4\times10^{-3} \le Y_{\rm{H}_2} \le  6\times10^{-3}$;
(f) $2\times10^{-3} \le Y_{\rm{H}_2} \le  3\times10^{-3}$;
(g) $1\times10^{-3} \le Y_{\rm{H}_2} \le  2\times10^{-3}$;
(i) $1\times10^{-4} \le Y_{\rm{H}_2} \le  5\times10^{-4}$.
}
\end{figure}
In order to distinguish the contribution of the different components of the 
strain rate, the top panel of figure~\ref{fig:6} addresses the joint pdf (jpdf) of 
$Y'_{\rm OH}$ and the deviatoric strain rate component aligned 
to the flame normal ${\bf n}$, 
$D_{nn}=\mathbf{E_{dev}}: \mathbf{n} \otimes  \mathbf{n}$, term II of equation~\eqref{stretch}.
Near the fresh gas, panels (b) and (c), a weak positive correlation is detected. Moving through the front 
towards the burnt gas, we observe an increasing positive correlation, see e.g. panel (d). 
In the middle of the flame
front, panels (f) and (g), the now familiar bi-modal behavior is retrieved together with a positive correlation 
between $Y_{{\rm OH}}'$ and $D_{nn} = 2/3\, \partial u_n/\partial n -
1/3\, \nabla_\pi \cdot \mathbf{u}_\pi$, where $\nabla_\pi$ 
is the gradient component tangent to the ideal flame surface.
The corresponding unstretched laminar flame, characterized by $D_{nn}^L = 2/3\, \partial 
u_n/\partial n$, is indicated for comparison by the black circles in the figure panels. 
The most probable states of the turbulent flame are either 
more active $Y_{\rm OH}' > 0$ or 
less active $Y_{\rm OH}' < 0$ than the laminar flame ($D_{nn} > D_{nn}^L$).

\begin{figure}[t!]
\centering
\includegraphics[width=.32\textwidth]{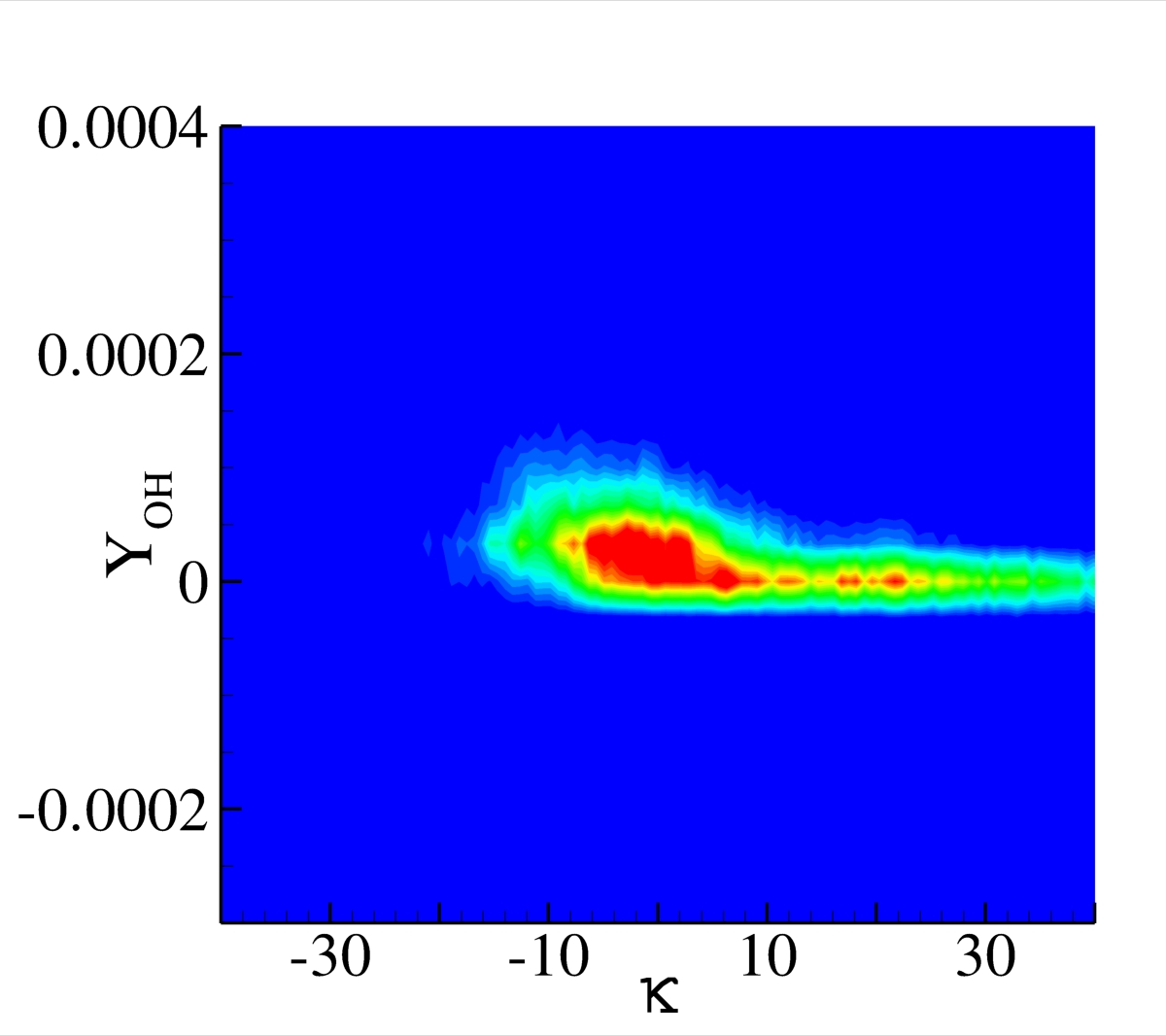}
\includegraphics[width=.32\textwidth]{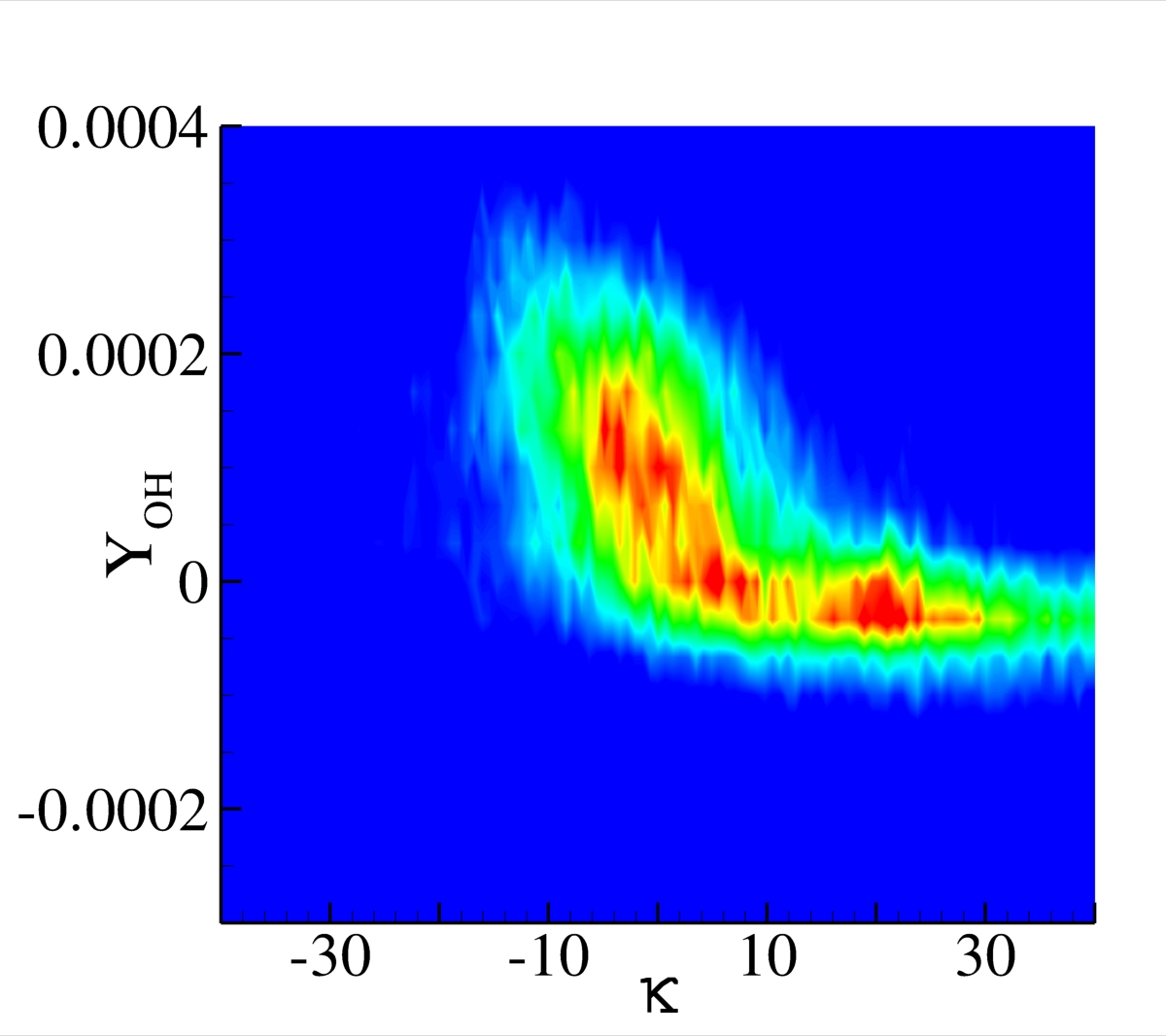}
\includegraphics[width=.32\textwidth]{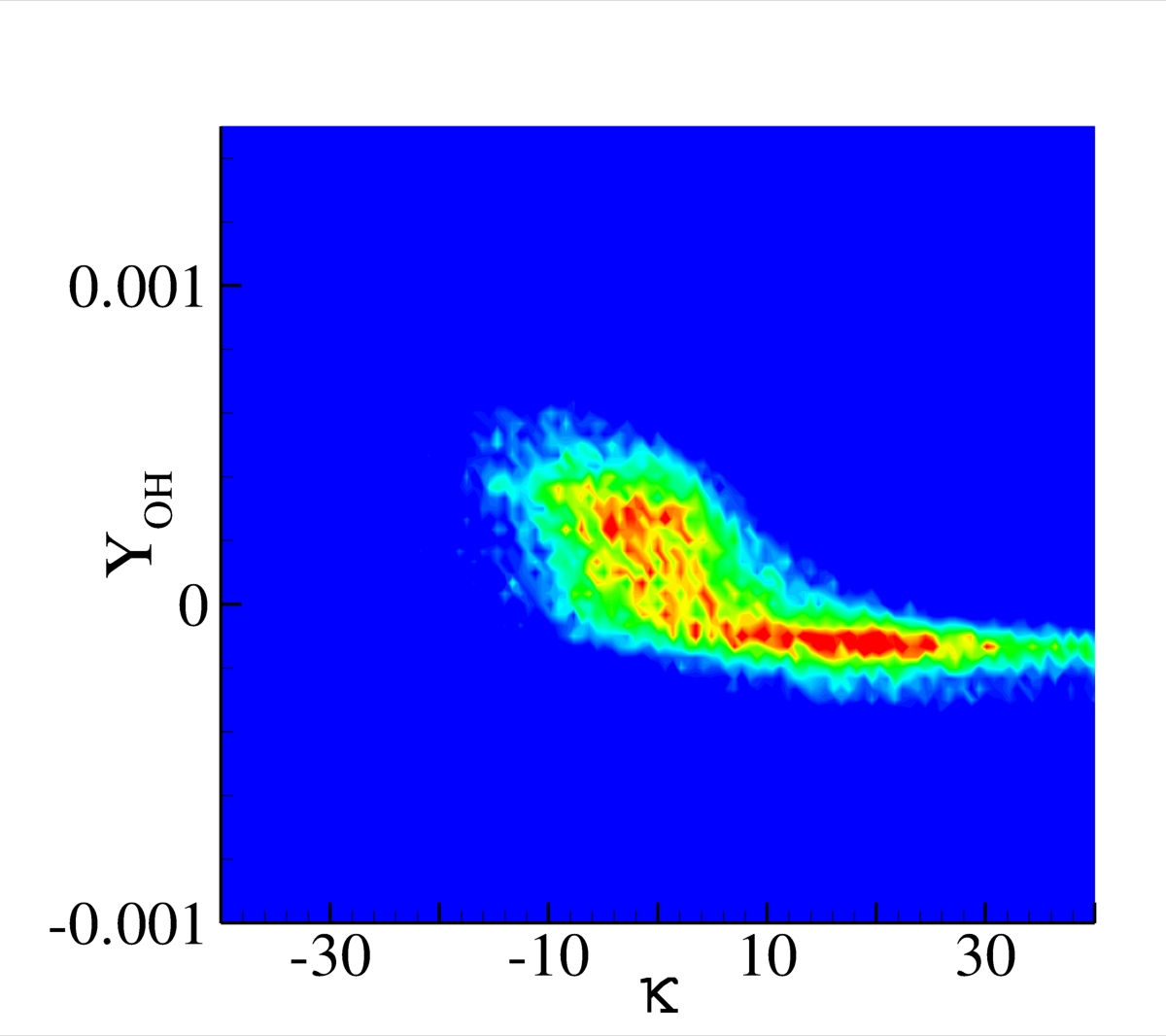}
{\scriptsize \put(-330,100){\bf (b)}}
{\scriptsize \put(-215,100){\bf (c)}}
{\scriptsize \put(-100,100){\bf (d)}}\\
\includegraphics[width=.32\textwidth]{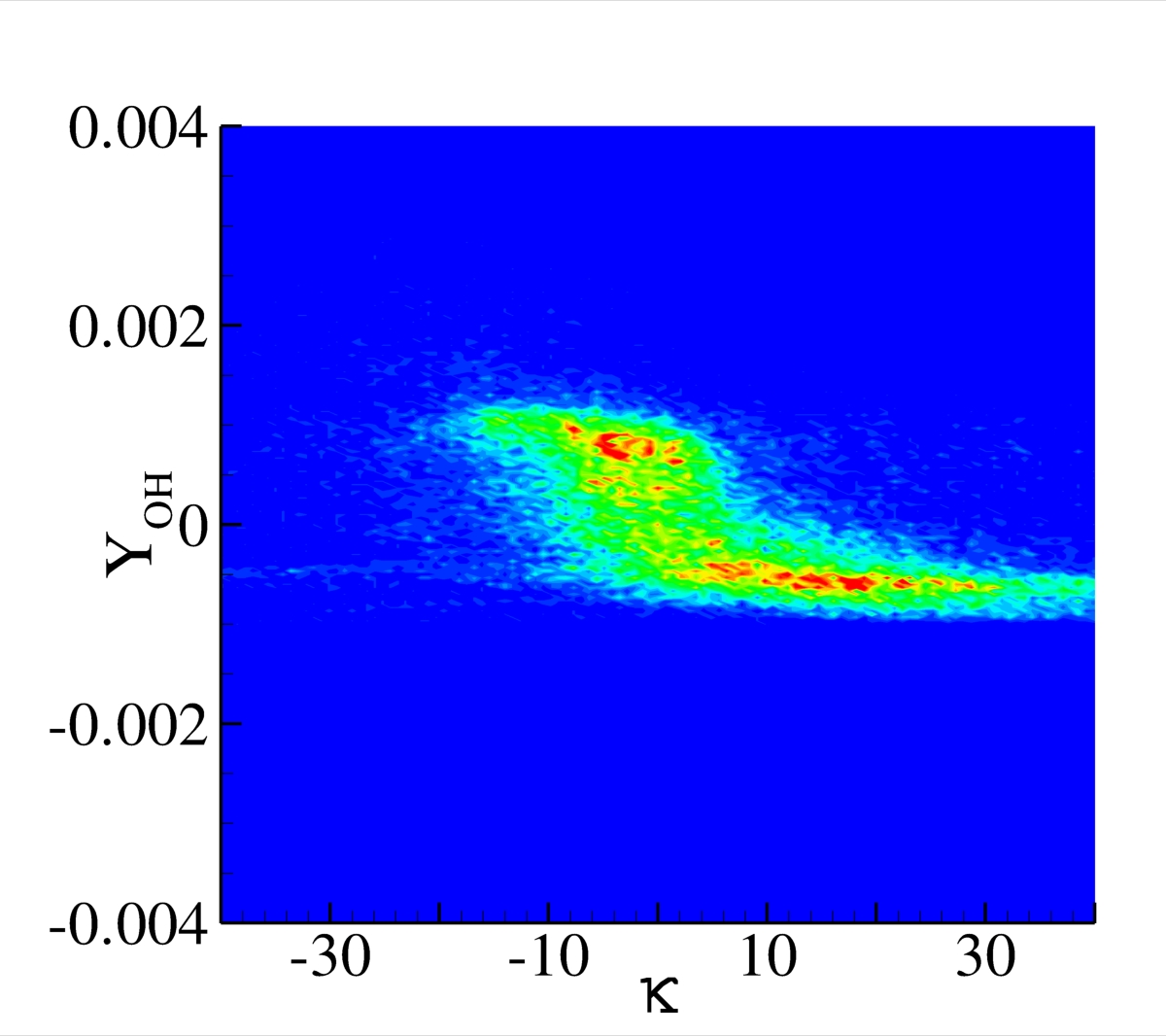}
\includegraphics[width=.32\textwidth]{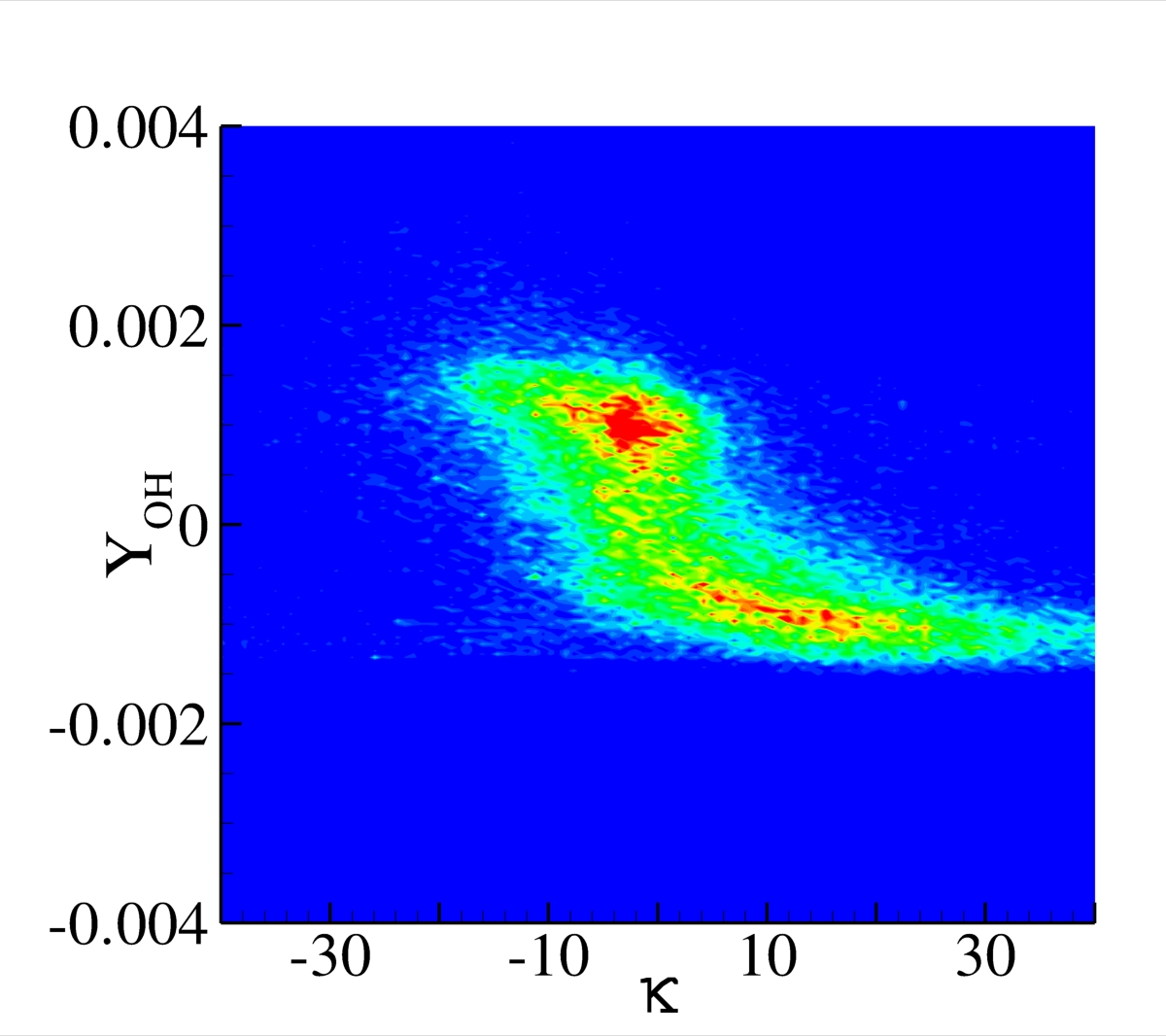}
\includegraphics[width=.32\textwidth]{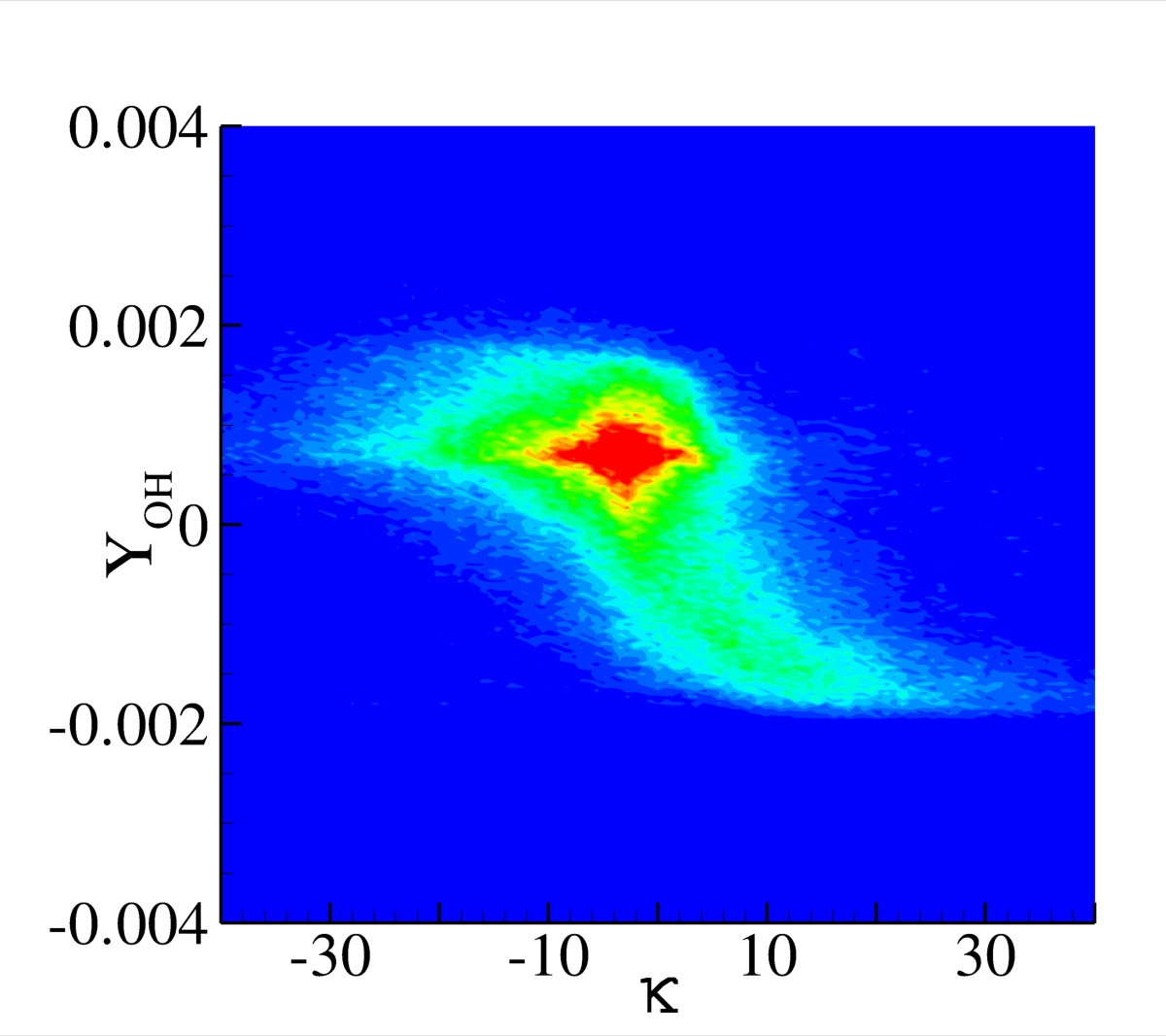}
{\scriptsize \put(-330,100){\bf (f)}}
{\scriptsize \put(-215,100){\bf (g)}}
{\scriptsize \put(-100,100){\bf (i)}}\\
\caption{\label{fig:7} Joint probability density function (jpdf) of
the flamelet normalized OH concentration $Y'_{\rm OH}$ and the local 
instantaneous curvature of the flame front evaluated by means as the 
$\nabla \cdot {\vec n}$ where ${\vec n}$ is the flame front normal 
${\vec n}=\nabla C_T/\left|\nabla C_T\right|$ with $C_T$ the temperature-based 
progress variable. In order to avoid spurious sampling effects, jpdf are evaluated along 
the flame front conditioning at different H$_2$ concentration intervals: 
(b) $8\times10^{-3} \le Y_{\rm{H}_2} \le  1\times10^{-2}$;
(c) $6\times10^{-3} \le Y_{\rm{H}_2} \le  8\times10^{-3}$;
(d) $4\times10^{-3} \le Y_{\rm{H}_2} \le  6\times10^{-3}$;
(f) $2\times10^{-3} \le Y_{\rm{H}_2} \le  3\times10^{-3}$;
(g) $1\times10^{-3} \le Y_{\rm{H}_2} \le  2\times10^{-3}$;
(i) $1\times10^{-4} \le Y_{\rm{H}_2} \le  5\times10^{-4}$.
 }
\end{figure}

A second relevant aspect is the effect of the local flame curvature,
\begin{equation}
k = \nabla \cdot \mathbf n =
-\left(\frac{1}{\mathcal{R}_1}+\frac{1}{\mathcal{R}_2}\right)
\end{equation}
where $\mathcal{R}_1$ and $\mathcal{R}_2$ are the principal curvature radii of the 
flame surface. 
The curvature helps defining the bulges (negative curvature) and 
gullies (positive curvature). Figure \ref{fig:7} shows the flame-conditioned jpdf 
of $Y'_{\rm OH}$ and mean curvature $k$. Near the fresh gases,
panel (b), the curvature has negligible effects on the OH radical concentration. Moving towards the burnt gas region, 
(c), (d), (f), (g), and (i) panels, the jpdf exhibits an evident negative correlation, confirming that  
strong burning states occur statistically in regions with prominent negative curvature (bulges), whereas 
quenching is typical of region with positive curvature (gullies).

\begin{figure}[h!]
\centering
\includegraphics[width=.32\textwidth]{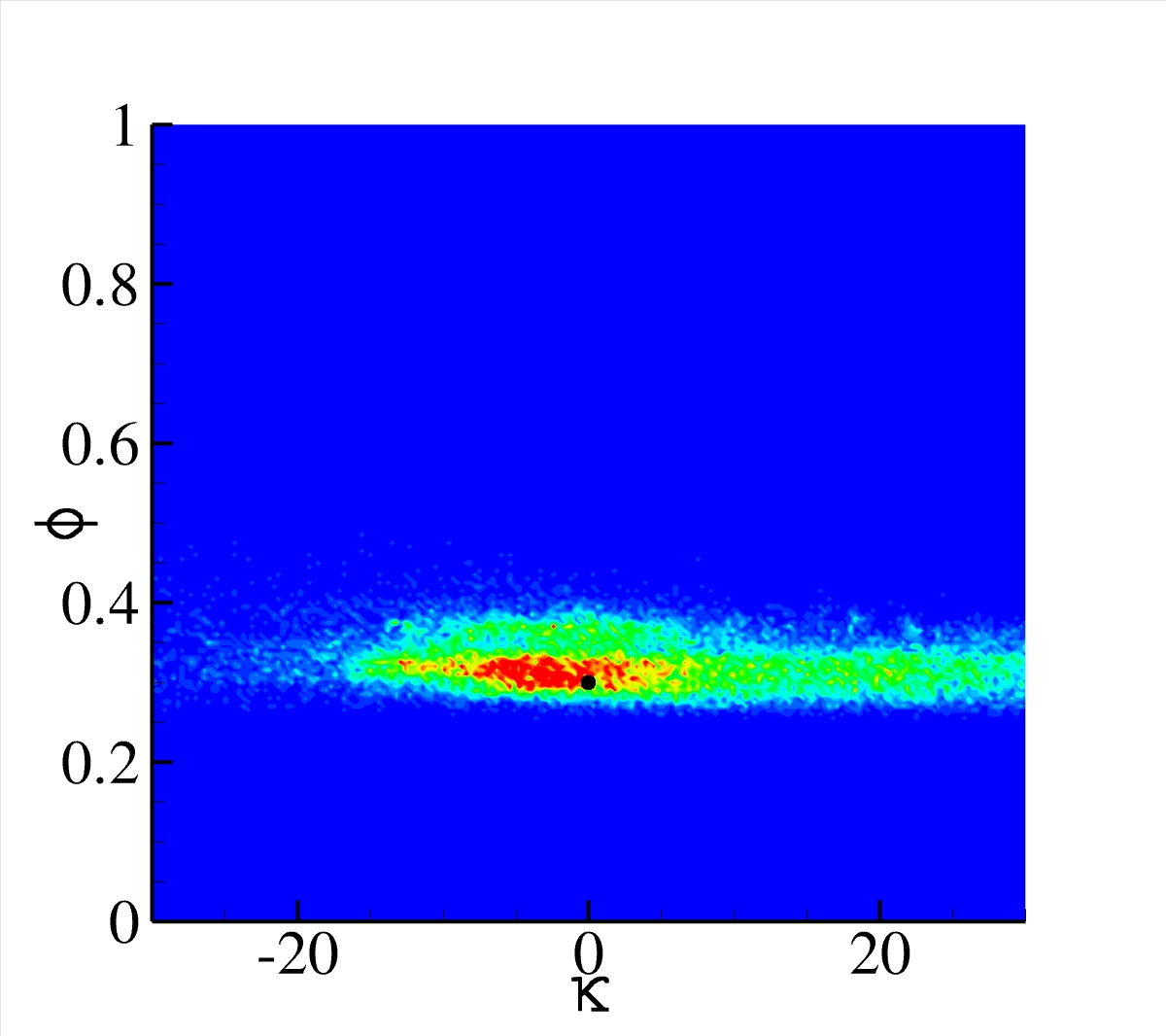}
\includegraphics[width=.32\textwidth]{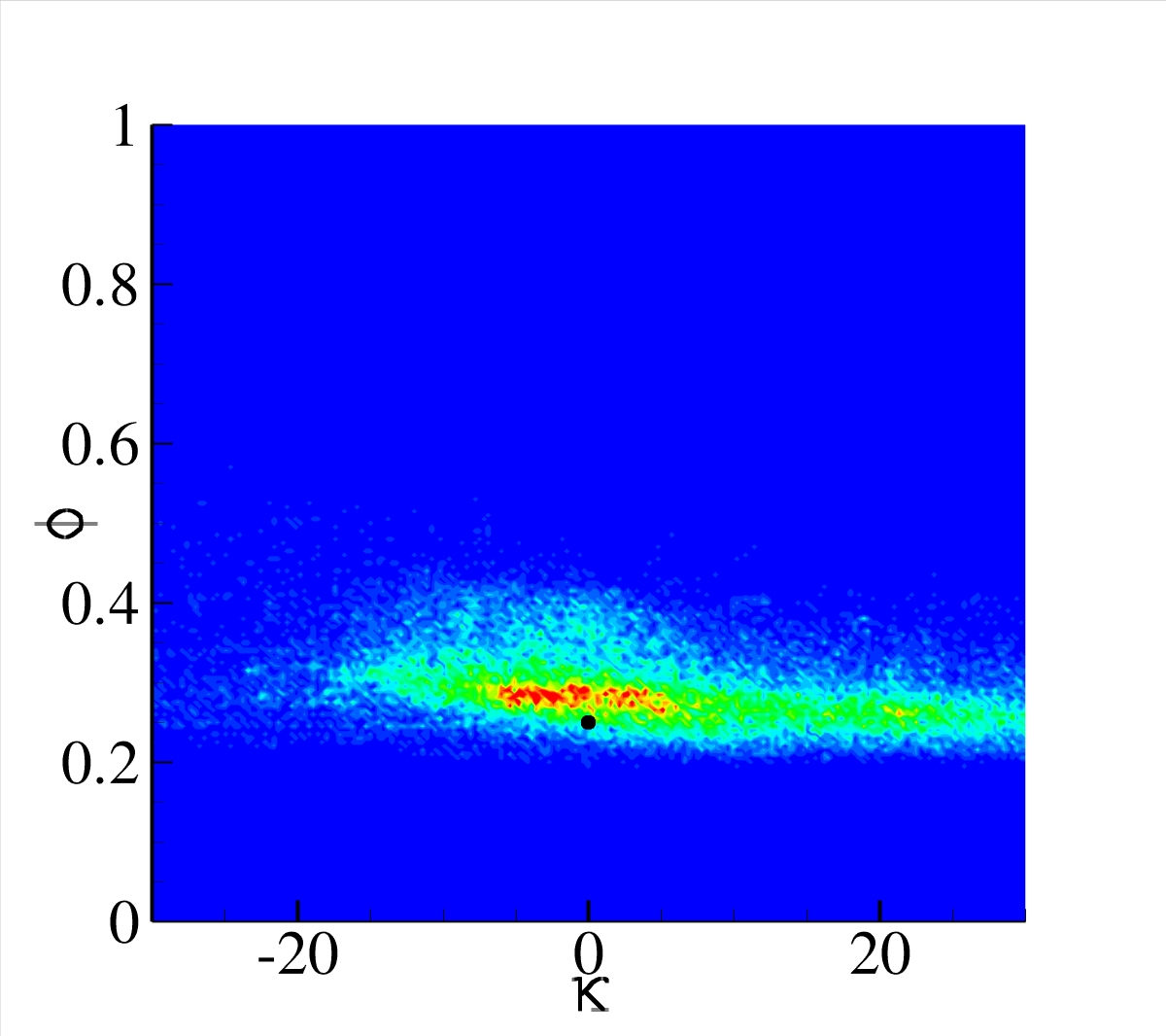}
\includegraphics[width=.32\textwidth]{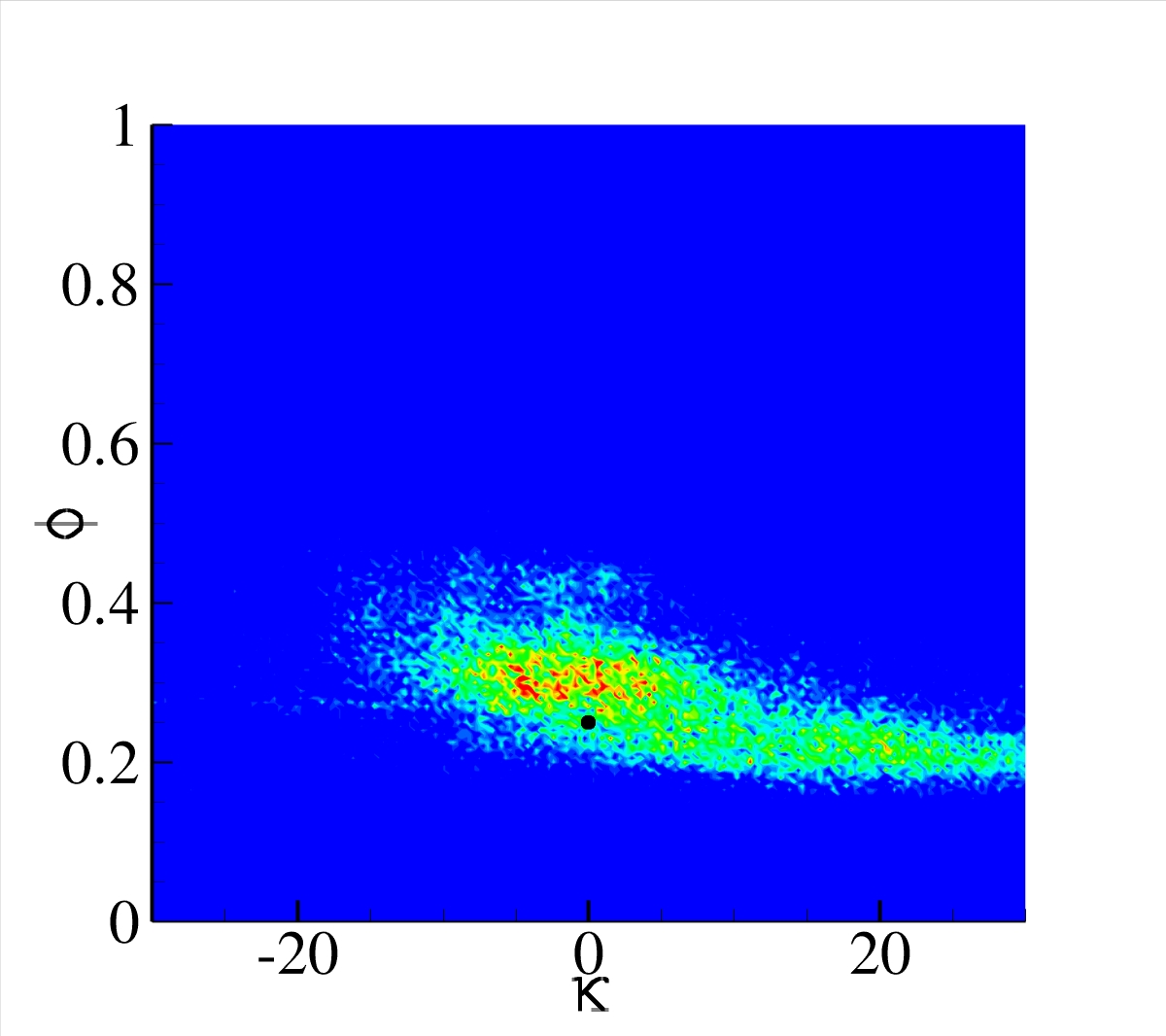}
{\scriptsize \put(-330,100){\bf (b)}}
{\scriptsize \put(-215,100){\bf (c)}}
{\scriptsize \put(-100,100){\bf (d)}}\\
\includegraphics[width=.32\textwidth]{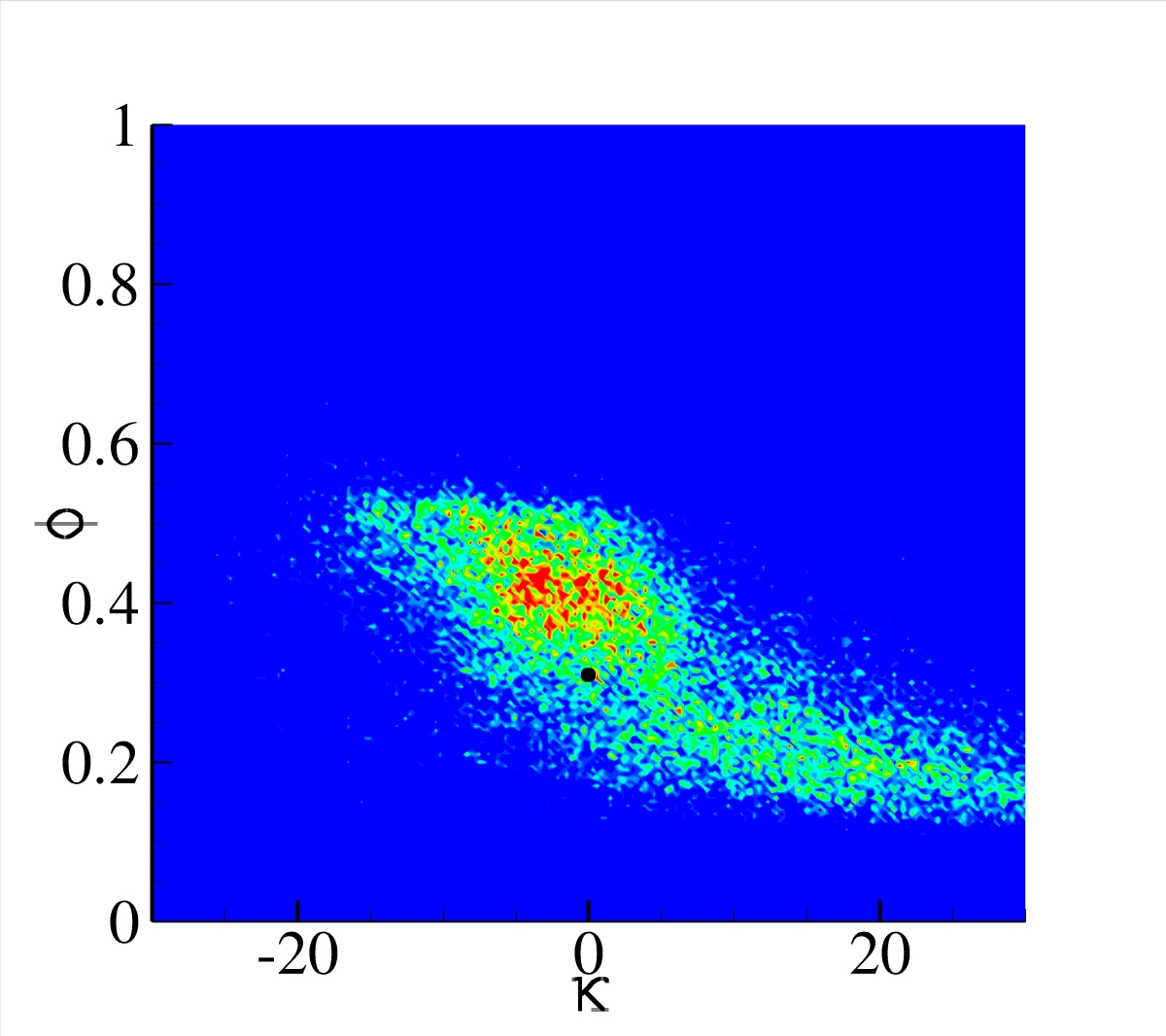}
\includegraphics[width=.32\textwidth]{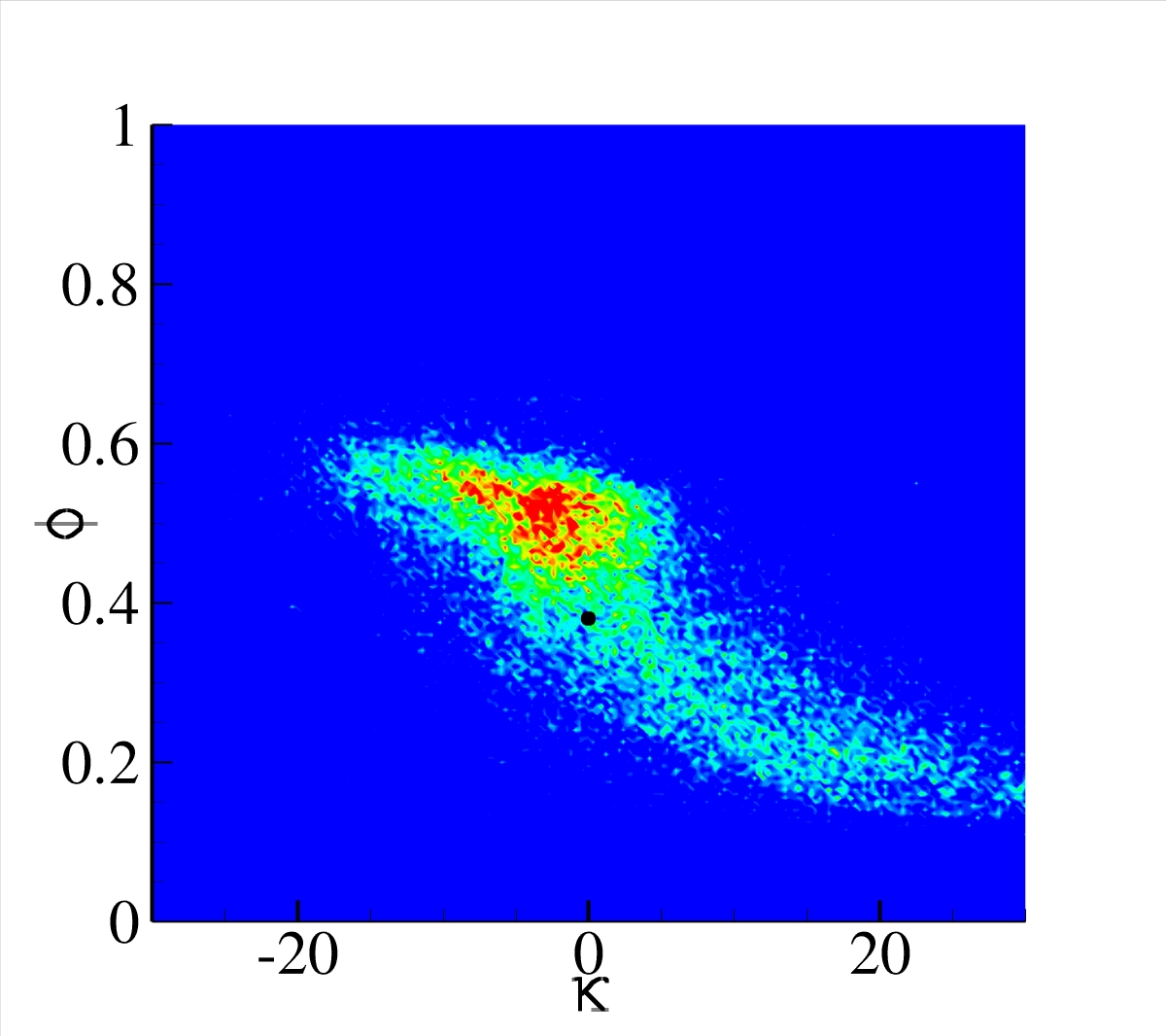}
\includegraphics[width=.32\textwidth]{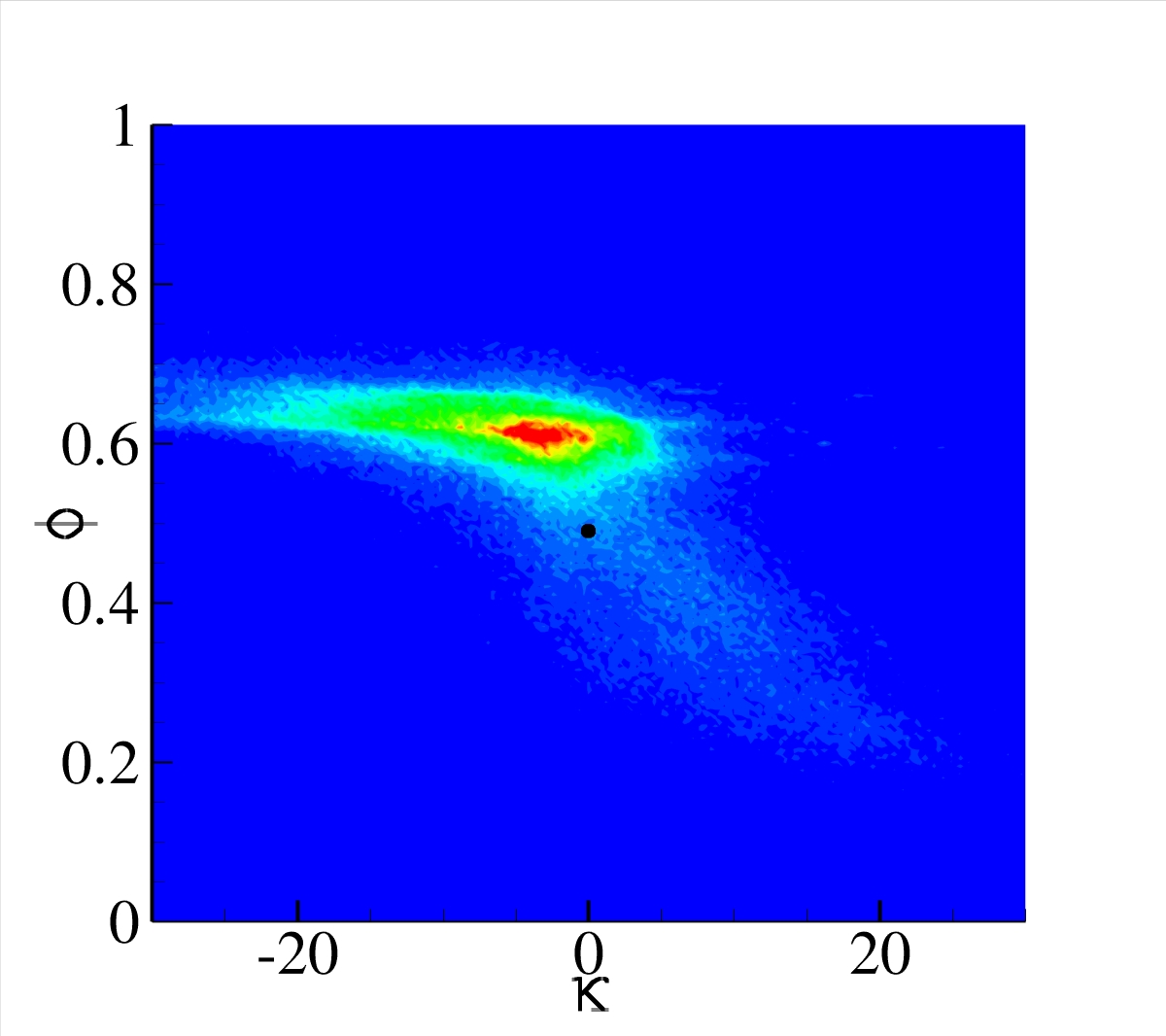}
{\scriptsize \put(-330,100){\bf (f)}}
{\scriptsize \put(-215,100){\bf (g)}}
{\scriptsize \put(-100,100){\bf (i)}}\\
\caption{\label{fig:8} Joint probability density function (jpdf) of the atomic equivalence 
ratio between H and O atoms: $\phi$, see eq.~\eqref{eq:phi_at} and the local 
instantaneous curvature of the flame front evaluated by means as the 
$\nabla \cdot {\vec n}$ where ${\vec n}$ is the flame front normal 
${\vec n}=\nabla C_T/\left|\nabla C_T\right|$ with $C_T$ the temperature-based 
progress variable. The black circle represents the unstretched laminar flame value. 
In order to avoid spurious sampling effects, jpdf are evaluated along 
the flame front conditioning at different H$_2$ concentration intervals: 
(b) $8\times10^{-3} \le Y_{\rm{H}_2} \le  1\times10^{-2}$;
(c) $6\times10^{-3} \le Y_{\rm{H}_2} \le  8\times10^{-3}$;
(d) $4\times10^{-3} \le Y_{\rm{H}_2} \le  6\times10^{-3}$;
(f) $2\times10^{-3} \le Y_{\rm{H}_2} \le  3\times10^{-3}$;
(g) $1\times10^{-3} \le Y_{\rm{H}_2} \le  2\times10^{-3}$;
(i) $1\times10^{-4} \le Y_{\rm{H}_2} \le  5\times10^{-4}$.
}
\end{figure}
As a further effect the flame curvature may also alter the local ratio between 
hydrogen and oxygen due to the different diffusion coefficients of the species. 
This is a well known in laminar lean hydrogen flames where thermo-diffusive 
instabilities occur. To measure the local equivalence ratio, it 
is instrumental to define the atomic equivalence ratio, see also~\cite{belchedayshe}. 
In lean flames the local molecular equivalence ratio decreases
up to vanish during the progress of combustion. To quantify the effect of the 
differential diffusion of the species, an equivalence ratio based
on the local number of H and O atoms is defined:
\begin{equation}
\label{eq:phi_at}
\phi=\frac{1}{2}\frac{X_{\rm H}+X_{\rm OH}+2\,X_{{\rm H}_2 {\rm O}}+2\,X_{{\rm H}_2}}{X_{\rm O}+X_{\rm OH}
+X_{\rm NO}+X_{{\rm H}_2 {\rm O}}+2\,X_{{\rm O}_2}},
\end{equation}
where $X_i$ denotes the molar fraction of the $i$-specie. This atomic 
equivalence ratio $\phi$ is constant during the whole reaction if the
species have the same diffusion coefficients. In the present case, a dilution
of the unburned  mixture may in principle due to entrainment of cold air from the environment.
This event however is quite rare since mixing usually takes place only
between the hot burnt gases and the cold ambient air.  
Figure~\ref{fig:8} shows the jpdf of $\phi$ and the mean curvature $k$ inside the
flame front for the same $Y_{{\rm H}_2}$ ranges of figure~\ref{fig:5}. 
At the beginning of the flame front, i.e.\ high $Y_{{\rm H}_2}$ (panels (b) and (c)),
the curvature has negligible effects on the atomic equivalence ratio as shown by the jpdf 
almost flat around the corresponding laminar unstretched value. It should be remarked that in laminar
hydrogen flames the atomic equivalence ratio is constant in the burnt and fresh gas regions.  However 
inside the flame front it diminishes, given the preferential 
diffusion of $\rm H_2$ which leads to a decrease of  hydrogen molar fraction, $X_{{\rm H}_2}$, and a slight increase of 
the oxygen molar fraction, $X_{{\rm O}_2}$,
(not shown) see e.g.\ pag. 287~\cite{law} where the phenomenon is accurately described. The corresponding 
laminar values in each conditioned jpdf is shown by a black circle. 
 In the middle
of the flame front, i.e. panels (d)-(f)-(g), the jpdfs show a strong 
anti-correlation between 
the curvature and the atomic equivalence ratio. 
Fronts with positive curvature (gullies) present low values of the atomic equivalence ratio, which is 
much smaller than the nominal value $\phi=0.5$ and the corresponding laminar one, 
leading to the quenched states already highlighted  in figure~\ref{fig:7}.  
On the contrary, fronts with negative curvature (bulges) exhibit 
enriched mixtures, that are responsible of the higher chemical activity and the super-adiabaticity as 
depicted in figure~\ref{fig:1}. Moreover, we note that the most probable atomic equivalence fraction is usually
higher than the corresponding laminar one (black circles) approaching the nominal value of $\phi=0.5$.
{In other words, it appears
that the turbulent fluctuations contribute to homogenize the mixture towards the nominal 
equivalence ratio, i.e.\ $\phi$ is often closer to the nominal bulk value than to the  
laminar unstrained flame value}. Approaching the end of the reaction region, panel (i), we note that the
most frequent state is that of the bulges with high atomic equivalence ratio.
{We remark that the departure of the atomic equivalence ratio from its nominal bulk value 
(the inlet one) occurs only within the flame while the nominal value is recovered ahead end behind the 
flame.}

{
\subsection{Turbulent vs Laminar Bunsen flame comparison}
\begin{figure}
\centering
\includegraphics[width=.75\textwidth]{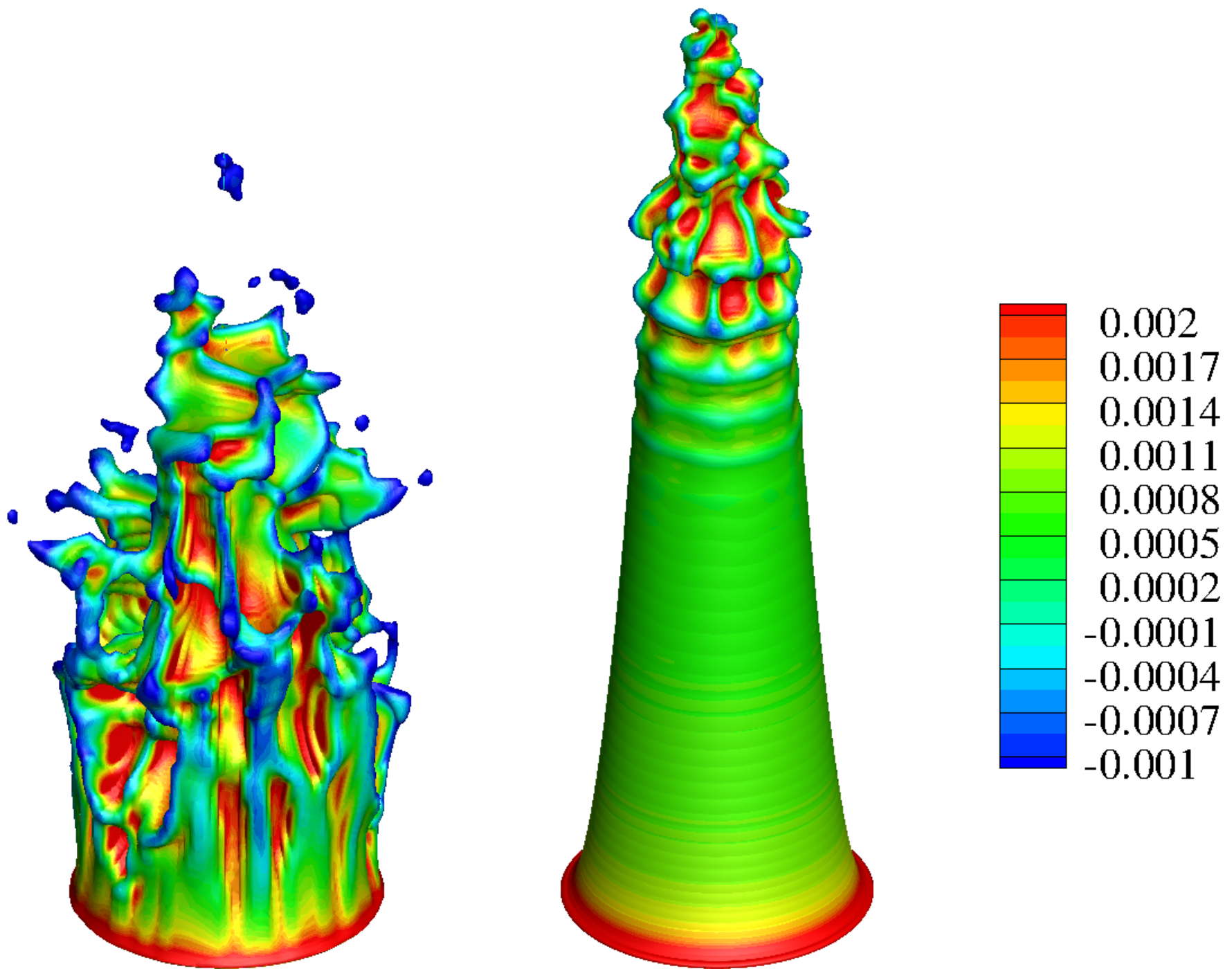}
\caption{\label{fig:lt_sur} Instantaneous isosurfaces of turbulent (left) and laminar (right)
inflow Bunsen flame. The isosurface corresponds to the $Y_{{\rm H}_2}=0.015$ isolevel and color map refers to
the value of the $Y'_{\rm OH}$, see eq.~\eqref{e:y'_oh}, in the range $0.01<Y_{{\rm H}_2}<0.02$, 
corresponding to the case (g) of figure~\ref{fig:5a}, which main feature is the bi-modal distribution of the $Y'_{\rm OH}$ 
pdf.}
\end{figure}
\begin{figure}
\centering
\includegraphics[width=.475\textwidth]{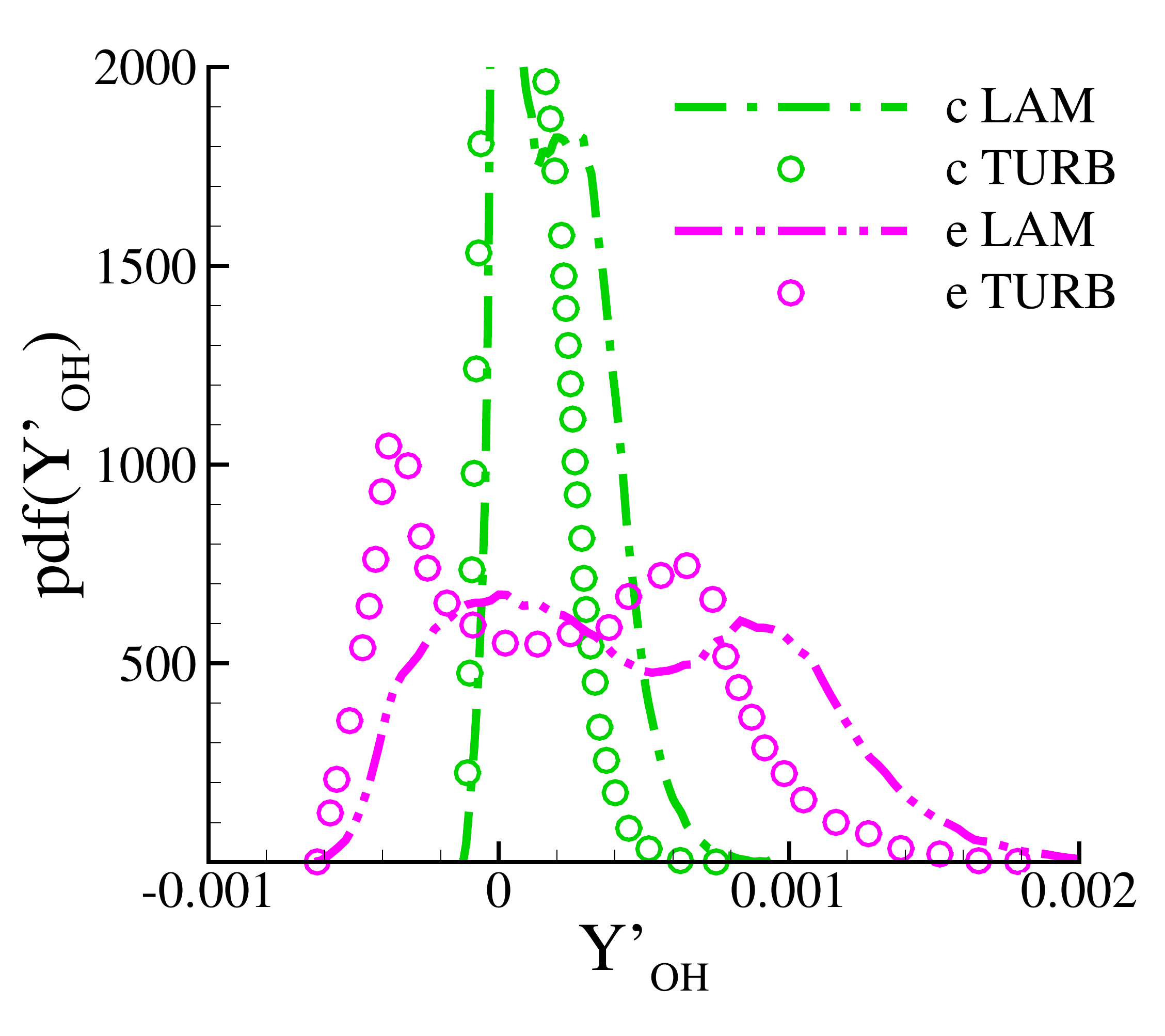}
\includegraphics[width=.475\textwidth]{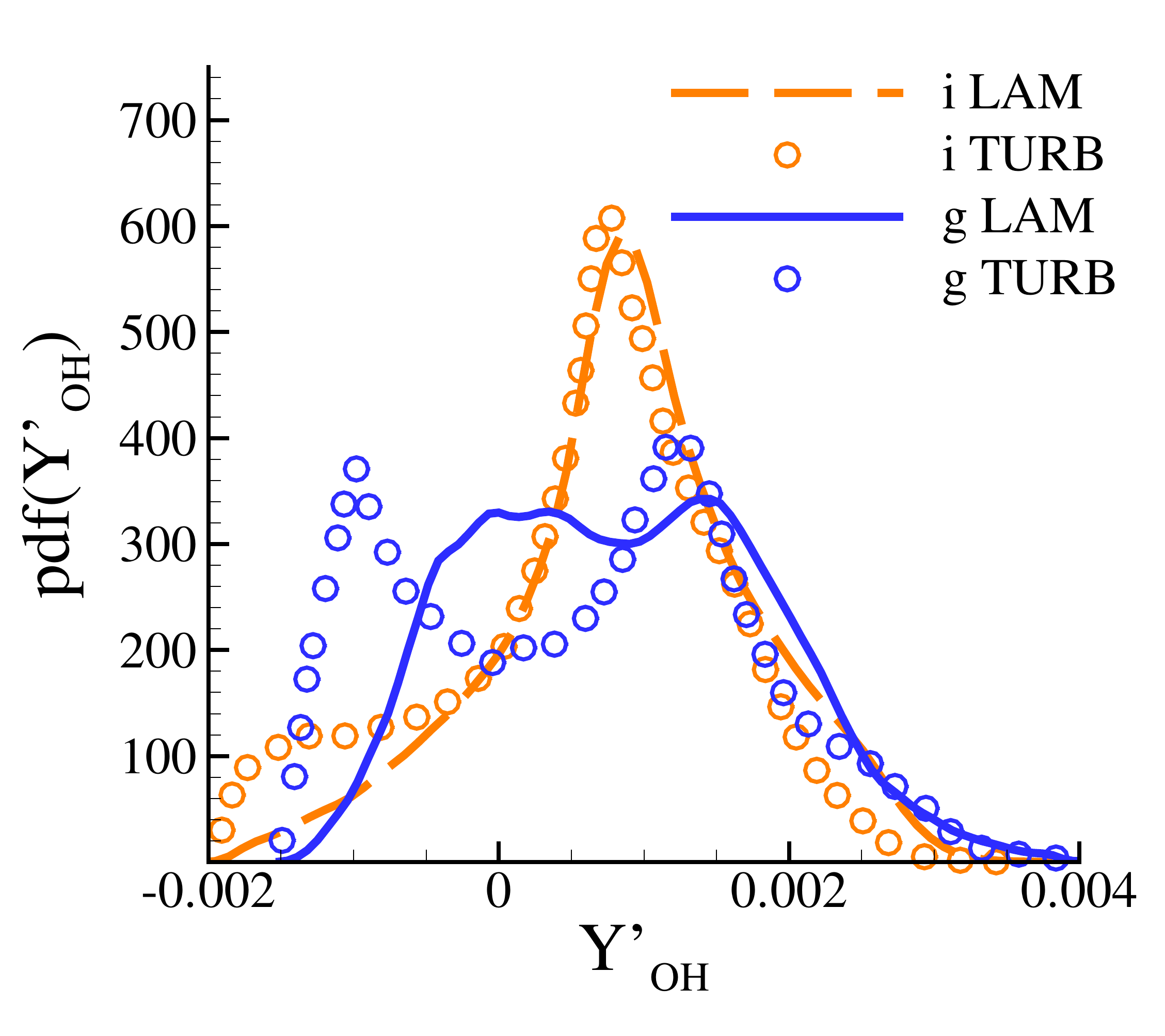}
\caption{\label{fig:OH_pdf_2} Probability density function of the OH radical mass flamelet fluctuation, see 
eq.~\eqref{fig:5a}, at 4 of the 8 different H$_2$ concentration intervals: 
(c) $6\times10^{-3} \le Y_{\rm{H}_2} \le  8\times10^{-3}$;
(e) $3\times10^{-3} \le Y_{\rm{H}_2} \le  4\times10^{-3}$;
(g) $1\times10^{-3} \le Y_{\rm{H}_2} \le  2\times10^{-3}$;
(i) $1\times10^{-4} \le Y_{\rm{H}_2} \le  5\times10^{-4}$.
Comparison between the laminar (lines) and the turbulent (symbols) Bunsen flame.
For the sake of clarity colors and line styles are kept the same of fig.~\ref{fig:5a}.
For the laminar inflow case, only the upper part of the flame, $5<z/R<7.5$,
(where thermo-diffusive instability 
is developed) has been considered for statistics.}
\end{figure}
}
{
In this subsection we provide a comparison between $H_2$/Air Bunsen flames fed by a steady 
laminar and the unsteady turbulent inflow in order to clarify  to which extent turbulence affects
 thermo-diffusive instability. The absence of noise in the laminar inflow prevents to observe transition
 to turbulence in the present case.\\
In Figure~\ref{fig:lt_sur}  instantaneous flame surfaces of the turbulent (left) and of the laminar (right) 
inflow Bunsen flames are reported. The surfaces are 
$Y_{{\rm H}_2}=0.015$ isolevels and are colored by OH radical 
flamelet fluctuation $Y'_{\rm OH}$. As expected, under turbulent inflow the flame is shorter 
indicating that, for a given flow rate, turbulent fluctuations enhance the overall burning speed. 
The turbulent flame surface is significantly wrinkled while the laminar one 
presents  corrugations only at tip and is smooth in the lower part.\\
Both flames show super-burning regions.} 
{
In the turbulent inflow case,
flame quenching cusps are present in the gullies, as denoted by the intense negative values of 
 $Y'_{\rm OH} < 0$, while in the laminar Bunsen flame cusps are almost absent, being 
$Y'_{\rm OH}\simeq 0$, i.e.\ the OH 
concentration is similar to the planar laminar case.\\}
{
%
Figure~\ref{fig:lt_sur} also illustrates the different paths followed by thermo-diffusive instabilities 
in laminar and turbulent  flames. 
In laminar Bunsen flames, the instability slowly evolves moving downstream. First   
perturbations originate and grow in planes normal to
the jet axis, then typical cellular-like structures form and eventually mutually
interact loosing their regularity near the flame tip.  
The turbulent case is strongly influenced by the
turbulent coherent structures issued by the turbulent pipe flow. 
In particular, the quasi-streamwise vortices together with the low- and high- speed streaks living near the pipe wall are much stronger than the small scale turbulence in the jet core and
(see e.g. \cite{robannrev,brandt2014lift}) 
control  the front curvature in the lower part of the flame, triggering corrugations 
parallel to the flow direction.
These corrugations generate super-burning and quenched regions via the thermo-diffusive 
instabilities.
Further downstream the front becomes more and more corrugated as expected of a
fully turbulent flow. 
\\}
{
The statistical characterization of these behaviors is discussed. 
For the two cases figure~\ref{fig:OH_pdf_2} provides the 
probability density function of $Y'_{\rm OH}$ conditioned to different H$_2$ concentration levels. 
In interval (c), i.e.\ near the fresh gas, fig.~\ref{fig:OH_pdf_2} left panel, a 
mono-modal behavior similar to the turbulent case is noted for the laminar Bunsen flame,
though with more frequent positive strong events than the turbulent case. 
In other words, the laminar flame shows relatively 
more regions with intense 
reaction rate and higher flamelet fluctuation of OH concentration than the turbulent one.
Different is the behavior well inside the flame front, e.g.\ in the intervals (e) and (g), 
where the turbulent inlet case shows the discussed bi-modal distribution which 
is either absent or strongly reduced in the laminar case. In particular, in interval (e), left panel 
of fig.~\ref{fig:OH_pdf_2}, the negative fluctuation peak of the turbulent case does not appear for
the laminar flame.
The pdf of $Y'_{\rm OH}$ for the laminar Bunsen flame is almost flat  between
zero and small positive values showing relatively more frequent strong super-burning regions.  
The conditioning interval (g), $1.\times 10^{-3} \le Y'_{H_2} \le 2. \times 10^{-3}$, correspond to
the isosurfaces shown in fig.~\ref{fig:lt_sur} ($Y'_{H_2} \le 1.5 \times 10^{-3}$). 
As observed, negative values of 
$Y'_{\rm OH}$ are less frequent than the turbulent case with an even  flatter pdf. 
A similar trend is found towards the end of the reaction front, interval (i), 
 where the probability to find negative OH flamelet fluctuations is smaller than the turbulent 
case. 
\\ 
In conclusion, though the thermo-diffusive instability crucially affects both turbulent and laminar flames, 
the presence of turbulent structures deeply alter its dynamics.
The laminar OH concentration pdf is always mono-modal with no negative peak. implying that 
quenched regions are statistically irrelevance.
Indeed turbulence is the source of the observed bimodal behavior, fig.~\ref{fig:5a}. 
The turbulent fluctuations produce corrugations of the flame front that in turn 
trigger the thermo-diffusive instability.  The 
probability of negative OH concentration flamelet fluctuations becomes significant, implying local quenching associated with positive curvature 
(gullies). Apparently, while super-burning regions are naturally developed by the thermo-diffusive
instability even in laminar flames, only the turbulence triggers local flame quenching. 
We have observed that the thermo-diffusive instability slowly grows from its inception 
at base up to flame tip where disturbances become three-dimensional.  
Turbulent fluctuations are instead much faster and sufficiently strong to prevent the formation 
of the thermo-diffusive cellular structures thereby setting the time scale of the turbulent flame.
}\\

\section{Final remarks}
\label{sec:concl}
The paper discusses DNS data of a turbulent Bunsen flame of a lean premixed 
mixture of molecular hydrogen and air. The turbulent combustion appears to be mainly characterized 
by two states: strong burning cells alternating with nearly quenched regions. Both the 
instantaneous behavior and the statistical investigation suggest that the 
strong burning cells are associated with the bulges, namely the regions convex towards the 
fresh gases (negative curvatures), while almost all quenched areas are associated 
with the gullies, regions convex towards the burnt gases (positive curvatures). 
To better characterize this behavior we have used a statistical analysis conditioned to the local
hydrogen concentration in order to move through the local reaction region from fresh to burnt gases.
Well inside the flame front,
the statistical occurrence of the bi-modal states exceeds the incidence of the corresponding laminar 
planar one, so that the flame may be considered  mainly constituted 
by these two states.  
A positive correlation is detected between the deviatoric strain rate and the local chemical activity;
moreover,  the expansion induced by the heat release 
produces a larger deformation in the direction normal to
the flame front, localized in the strong burning cells.
The analysis of the front curvature  suggests 
that the origin of such bi-stable behavior is related to the thermo-diffusive 
instabilities. The local curvature of the front, combined with the different transport 
properties of the species, induces a local enrichment/dilution of the mixture 
and it results in strong burning cells (bulges) and quenched regions (gullies). 
The local curvature was shown to correlate with the local equivalence
ratio and produces the observed bimodal behavior. 
In addition, the enrichment/depletion of the local mixture is spatially 
alternated along the flame front.
Since the bulk equivalence ratio of the mixture was fixed to be $\Phi = 0.5$, 
the local enrichment and depletion  of the mixture correspond to an 
increase/decrease of the adiabatic combustion temperature. 
The picture emerging from the present moderate Reynolds number simulation is consistent with 
the turbulent fluctuations initially corrugating the flame front and triggering the instabilities which successively 
affect the local chemical activity.

Besides the regions well inside reaction zone, 
in the flame region closest to the fresh and burn gases,  
a mono-modal behavior similar to the unstretched laminar flame is 
detected. However, the  bi-stable dynamics influences the process 
also in these external parts of the flame as indicated by the strong pdf tails of the flamelet OH concentration.

{A comparison between the turbulent and a laminar inflow Bunsen 
flame was able to better assess the effect of turbulent fluctuations. 
The laminar flame shows a clear cellular shape near the tip, purely induced by the 
thermo-diffusive instability. Differently from the turbulent flame, the laminar flame 
process is alway mono-modal. Actually we found a high statistical incidence 
of the super-burning state, but not locally quenched regions. We infer that the super-burning behavior
naturally develops via thermo-diffusive instability even without turbulent fluctuations, which are instead necessary to
 originate quenched regions triggering the formation of gullies.}

These  results are expected to be relevant in the context of modeling  
turbulent flames at low Lewis numbers by means of the flamelet assumption, 
since the local flame features are frequently and markedly different
from the unstretched laminar flame. In particular, flame models should
 comply with the bi-stable nature of the 
turbulent flame which is mainly 
characterized  by two alternative local states, the super-adiabatic and the almost quenched state.

\bibliographystyle{spphys}       
\bibliography{biblio}   


\end{document}